\documentclass[12pt]{article}
\usepackage{epsf,amsfonts,amssymb,epsfig,amsmath,mathtools,graphics,slashed}
\usepackage{hyperref,hep,graphicx,subfig}
\usepackage{rotating}

\newcommand{\nn}{\notag \\}

\DeclareMathOperator{\tr}{Tr}
\def\vev#1{\langle\, #1 \, \rangle}

\makeatletter

\@addtoreset{equation}{section}
\makeatother

\begin{document}

\begin{titlepage}

\vfill

\begin{flushright}
Imperial/TP/2017/JG/04\\
DCPT-17/21
\end{flushright}

\vfill

\begin{center}
   \baselineskip=16pt
   {\Large\bf Diffusion in 
   inhomogeneous media}
  \vskip 1.5cm
  \vskip 1.5cm
Aristomenis Donos$^1$, Jerome P. Gauntlett$^2$ and Vaios Ziogas$^1$\\
     \vskip .6cm
     \begin{small}
      \textit{$^1$Centre for Particle Theory and Department of Mathematical Sciences\\Durham University, Durham, DH1 3LE, U.K.}
        \end{small}\\    
         \begin{small}\vskip .6cm
      \textit{$^2$Blackett Laboratory, 
  Imperial College\\ London, SW7 2AZ, U.K.}
        \end{small}\\
        \end{center}
     \vskip .6cm
\vfill

\begin{center}
\textbf{Abstract}
\end{center}
\begin{quote}
We consider the transport of conserved charges in spatially inhomogeneous quantum systems with a discrete lattice symmetry.
We analyse the retarded two point functions involving the charges and the associated currents at long wavelengths, compared to the scale of the lattice, and, when the DC conductivities are finite, extract the hydrodynamic modes associated with diffusion of the charges.
We show that the dispersion relations of these modes are related to the eigenvalues of a specific matrix constructed from
the DC conductivities and certain thermodynamic susceptibilities, thus obtaining generalised Einstein relations.
We illustrate these general results in the specific context of relativistic hydrodynamics 
where translation invariance is broken using spatially inhomogeneous and periodic deformations
of the stress tensor and the conserved $U(1)$ currents. Equivalently, this corresponds to considering
hydrodynamics on a curved manifold, with a spatially periodic metric and chemical potential, and we obtain the dispersion
relations for the heat and charge diffusive modes.
\end{quote}

\vfill

\end{titlepage}

\setcounter{equation}{0}
\section{Introduction}

Motivated by various strongly correlated states of matter seen in Nature, there has been a significant effort devoted to 
obtaining a deeper theoretical understanding of thermoelectric transport. It has long been appreciated that it is necessary to 
work within a framework in which momentum is not conserved. Indeed, for a translationally invariant system in which momentum is exactly conserved, 
the AC thermal response necessarily contains a delta function at zero frequency leading to a non-physical infinite DC thermal conductivity. 
Thus, one is interested in setups in which translation symmetry is explicitly broken. 

In this paper we will present some general results for thermoelectric transport in inhomogeneous systems. More precisely, we will consider
general quantum systems, with one or more conserved currents, with a discrete, spatial lattice symmetry. This could describe, for example, 
a quantum field theory in which translation invariance has been explicitly broken by deforming the theory with operators which have a periodic dependence on the spatial coordinates.

Of central interest are the retarded Green's functions for the current-current correlators $G_{JJ}(t,{\bf x};t'{\bf x}')$. At the level of linear response
these determine how the currents respond after perturbing the system by a current source. Time translation invariance implies
that these Green's functions only depend on $t-t'$ which allows us to Fourier transform and obtain $G_{JJ}(\omega, {\bf x},{\bf x}')$. 
In a translationally invariant setting the Green's functions would also only depend on ${\bf x}'-{\bf x}$ and a Fourier transform leads 
to a correlator depending on $\omega$ and a single wave-vector ${\bf k}$. 
When translations are broken, this is no longer possible but a discrete lattice symmetry allows us to define an infinite discrete set of correlators $G_{JJ}^{(\{n_i\})}(\omega, {\bf k})$, where $\{n_i\}$ are a set of integers. We will be particularly
interested in studying the correlator $G_{JJ}(\omega, {\bf k})\equiv G_{JJ}^{(\{0\})}(\omega, {\bf k})$. Indeed this correlator,
which satisfies a simple positivity condition, 
captures the transport properties of the system at late times and for wavelengths much longer than the scale of the lattice,
and thus we might call $G_{JJ}(\omega, {\bf k})$ a `hydrodynamic-mode correlator'.

By generalising similar computations presented in \cite{Forster} in the translationally invariant setting,
we will show that when the thermoelectric DC 
conductivity is finite, subject to some analyticity assumptions, 
there is necessarily a diffusion pole in the hydrodynamic mode correlator for $\omega,\varepsilon {\bf k}\to 0$. 
If the system has just a single conserved current then we explain precisely when we get a dispersion relation for the diffusion pole of the form 
\begin{align}\label{feinrint}
\omega=-i\varepsilon^2D({\bf k})+\dots\,, \qquad D({\bf k})=[\sigma^{ij}_{DC}k_ik_j]\chi({\bf 0})^{-1}\,,
\end{align}
where $\sigma^{ij}_{DC}$ is the DC conductivity and $\chi({\bf k})$ is the charge susceptibility. This is our first Einstein relation for inhomogeneous
media.

When there are additional conserved currents, there will be additional diffusion modes when the associated DC conductivities are finite.
We analyse the dispersion relations for the diffusion modes and show how they can be obtained from the eigenvalues of
a specific `generalised diffusion matrix' that is constructed from the DC conductivities and various thermodynamic susceptibilities.
We emphasise that, generically, the dispersion relations for the diffusion modes are not of the form \eqref{feinrint} and hence we 
refer to our result concerning the dispersion relation as a `generalised Einstein relation'. This feature of diffusion modes was also emphasised in
\cite{Hartnoll:2014lpa} within a specific hydrodynamic setting, which we will return to later.

These results concerning hydrodynamic modes of the Green's functions are very general.
However, motivated by recent experimental progress
\cite{2016Sci...351.1055B,2016Sci...351.1058C,2016Sci...351.1061M}, there has been considerable theoretical work using hydrodynamics to study thermoelectric transport \cite{Hartnoll:2014lpa,Davison:2014lua,Blake:2015epa,Lucas:2015lna,Davison:2015taa,Lucas:2015sya,Davison:2016hno,2017PNAS..114.3068G,Lucas:2016yfl,Banks:2016wdh,Delacretaz:2016ivq,Delacretaz:2017zxd,Lucas:2017ggp,Scopelliti:2017sga} and it is therefore of 
interest to see how our general results on diffusion manifest themselves in this particular context. More specifically, we will study this within the context of
relativistic hydrodynamics, describing the hydrodynamic limit of a relativistic quantum field theory.

Within this hydrodynamic framework, we first need to consider how momentum dissipation is to be incorporated. 
A standard approach is to modify, by hand, the hydrodynamic equations of motion, 
i.e. the Ward identities of the underlying field theory, 
by a phenomenological term that incorporates momentum dissipation (e.g. \cite{1963AnPhy..24..419K,Hartnoll:2007ih}). An alternative and more controlled approach is to maintain the Ward identities, which are fundamental properties of the field theory, but to consider the field theory to 
be deformed by spatially dependent sources. 
In this spirit, the hydrodynamic limit of a class of field theories which have been deformed by certain scalar operators was analysed in \cite{Blake:2015epa}. Subsequently, the universal class of deformations which involve adding spatially dependent
sources for the stress tensor were studied in \cite{Banks:2016wdh}. 
Since the stress tensor of the field theory couples to the spacetime metric, the deformations studied
in \cite{Banks:2016wdh} are equivalent
to studying the hydrodynamic limit of the quantum field theory on a curved spacetime manifold. The spacetime metric is taken to have a time-like Killing vector in order to discuss thermal equilibrium. Then, while spatial momentum will, generically, no longer be conserved, energy still will be. It may be possible to experimentally realise
the deformations studied in \cite{Banks:2016wdh} in real materials, such as strained graphene
\cite{RevModPhys.81.109,PhysRevB.81.035411,C5NR07755A}.

In this paper we extend the analysis of \cite{Banks:2016wdh} to cover relativistic quantum field theories which have
a conserved $U(1)$ symmetry\footnote{While writing up this work, ref. \cite{Scopelliti:2017sga} appeared which
also generalises \cite{Banks:2016wdh} to include a conserved $U(1)$ charge
and independently derived the hydrodynamic equations \eqref{testokes}, for the special case of no time dependence 
and for curved manifolds with a unit norm timelike Killing vector (i.e. $f=1$).}. As in \cite{Banks:2016wdh} we can consider the field theory to live on a static, 
curved manifold. Although not necessary, it will be convenient to take the manifold to have planar topology
and with a metric that is periodic in the spatial directions.
Within the hydrodynamic framework we will also consider deformations 
that are associated with spatially dependent sources for the $U(1)$ symmetry. This is particularly
interesting since it corresponds to allowing for spatially dependent chemical potential or, equivalently, spatially dependent charge density. One can anticipate that our results will be useful for understanding thermoelectric transport in real systems, such as charged puddles, with or without strain, in graphene \cite{2008NatPh...4..144M,2008PhRvL.101p6803R,2009NatPh...5..722Z,2011NatMa..10..282X} as also discussed in \cite{Scopelliti:2017sga}. 

As an application of our formalism, we show how to construct long-wavelength, late-time hydrodynamic modes that are associated with
diffusion of both energy and electric charge. We derive the dispersion relation for these modes and explicitly obtain the generalised
Einstein relations. It is worth noting that this result is independent of the precise transport coefficients that enter the constitutive relations in the conserved currents. 
We also note that a derivation of an Einstein relation for the diffusion of electric charge in the context of hydrodynamics
with vanishing local charge density in one spatial direction was carried out in appendix A of 
\cite{Lucas:2016yfl} and this is consistent with our more general analysis here.

\section{Green's function perspective}\label{green}

We begin our discussion with a general quantum system with a time independent Hamiltonian $H$. We assume that there
is a lattice symmetry group which acts on the $d$ spatial coordinates via $\mathbf{x}\to \mathbf{x}+\mathbf{L}_{j}$ and $U^{-1}_{\mathbf{L}_{j}}A(t,\mathbf{x})U_{\mathbf{L}_{j}}=A(t,\mathbf{x}+\mathbf{L}_{j})$, where $A(t,{\bf x})$ is an arbitrary local operator.
We assume that the Hamiltonian is invariant under this symmetry and hence $U^{-1}_{\mathbf{L}_{j}}HU_{\mathbf{L}_{j}}=H$. 
We will also consider the system to be at finite temperature $T$.

As usual, for two local operators $A(t,{\bf x})$, $B(t,{\bf x})$, the retarded two point functions are defined through
\begin{align}
G_{AB}(t,\mathbf{x}; t^{\prime},\mathbf{x}^{\prime})=-i\theta(t-t^{\prime})\,\vev{\left[ A(t,\mathbf{x}), B(t^{\prime},\mathbf{x}^{\prime})\right] }\,,
\end{align}
with $\vev{ A(t,\mathbf{x})}=\tr \left( \rho A(t,\mathbf{x}) \right)$, where $\rho=e^{-\beta H}/\tr (e^{-\beta H})$ and $\beta=1/T$. 
Using the fact that $A(t,\mathbf{x})=e^{i t H}A(0,\mathbf{x})e^{-i t H}$ and the lattice symmetry of $H$, we see that the two point functions will satisfy
\begin{align}
G_{AB}(t,\mathbf{x}; t^{\prime},\mathbf{x}^{\prime})&=G_{AB}(t-t^{\prime},\mathbf{x}; 0,\mathbf{x}^{\prime})\,,\label{eq:tsym}\\
G_{AB}(t,\mathbf{x}+\mathbf{L}_{j}; t^{\prime},\mathbf{x}^{\prime}+\mathbf{L}_{j})&=G_{AB}(t,\mathbf{x}; t^{\prime},\mathbf{x}^{\prime})\label{eq:lsym}\,.
\end{align}
The symmetry \eqref{eq:tsym} allows us to define a function with three arguments through $G_{AB}(t-t^{\prime},\mathbf{x},\mathbf{x}^{\prime})\equiv G_{AB}(t,\mathbf{x}; t^{\prime},\mathbf{x}^{\prime})$.

We next recall that if we introduce a perturbative source term in the Hamiltonian via 
\begin{align}\label{delaich}
\delta{H}(t)&=
\int d\mathbf{x}\, \delta h_{B}(t,\mathbf{x}){B}(t,\mathbf{x})\,,
\end{align}
then at the level of linear response, the change in the expectation values of an arbitrary operator ${A}$ is given by
\begin{align}\label{eq:response}
\delta\vev{{A}}(t,\mathbf{x})&=\int dt^{\prime}d\mathbf{x}^{\prime}\, G_{AB}(t-t^{\prime},\mathbf{x},\mathbf{x}^{\prime}) \delta h_{B}(t^\prime,\mathbf{x}^{\prime})\,.
\end{align}
We note that the source, and hence the response, need not be a periodic function of the spatial coordinates and indeed
this will be case of most interest in the following.

To proceed we Fourier transform the Green's function on all arguments and define
\begin{align}\label{eq:g_fourier}
G_{AB}(\omega,\mathbf{k},\mathbf{k}^{\prime})&\equiv \int dt d\mathbf{x}d\mathbf{x}^{\prime}\,e^{i\omega t-i\mathbf{k}\mathbf{x}+i\,\mathbf{k}^{\prime}\mathbf{x}^{\prime}}\,G_{AB} (t,\mathbf{x},\mathbf{x}^{\prime})\,.
\end{align}
The discrete symmetry \eqref{eq:lsym} implies that we can perform a crystallographic type of decomposition to obtain
\begin{align}\label{eq:g_cystal}
G_{AB}(\omega,\mathbf{k},\mathbf{k}^{\prime})=\sum_{\{n_{j}\}} G^{(\{n_{j}\})}_{AB}(\omega,\mathbf{k}^{\prime})\,\delta(\mathbf{k}-\mathbf{k}^\prime-n_{j}\,\mathbf{k}_{L}^{j}) \,,
\end{align}
where $\mathbf{k}_{L}^{j}$ are the reciprocal lattice vectors satisfying $\mathbf{k}_{L}^{i}\cdot\mathbf{L}^{j}=2\pi\delta^{ij}$ and $\{n_{j}\}$ are sets of integers. To see this, we simply notice that if we define the function
\begin{align}
G_{AB}(\omega,\mathbf{x},\mathbf{k}^{\prime})\equiv \int dt d\mathbf{x}^{\prime}\,e^{i\omega t+i\,\mathbf{k}^{\prime}\mathbf{x}^{\prime}}\,G_{AB} (t,\mathbf{x},\mathbf{x}^{\prime})\,,
\end{align}
then the real space lattice symmetry \eqref{eq:lsym} implies the periodicity condition
\begin{align}
G_{AB}(\omega,\mathbf{x}+\mathbf{L}_{j},\mathbf{k}^{\prime})=e^{i\,\mathbf{k}^{\prime}\mathbf{L}_{j}}\,G_{AB}(\omega,\mathbf{x},\mathbf{k}^{\prime})\,,
\end{align}
and hence we can deduce that $e^{-i\,\mathbf{k}^{\prime}\mathbf{x}}\,G_{BB^{\prime}}(\omega,\mathbf{x},\mathbf{k}^{\prime})$ is periodic as a function of $\mathbf{x}$. This lets us write it as a discrete Fourier series, expressing
\begin{align}\label{GtoGn}
G_{AB}(\omega,\mathbf{x},\mathbf{k}^{\prime})=\frac{1}{(2\pi)^{d}}\,e^{i\,\mathbf{k}^{\prime}\mathbf{x}}\,
\sum_{\{n_{j}\}}e^{i n_{j}\mathbf{k}^{j}_{L}\mathbf{x}} G^{(\{n_{j}\})}_{AB}(\omega,\mathbf{k}^{\prime})\,,
\end{align}
and \eqref{eq:g_cystal} follows.

In the sequel, we will be particularly interested in the zero modes, $G_{AB}(\omega,\mathbf{k})\equiv G^{(\{0\})}_{AB}(\omega,\mathbf{k})$.
These can easily be obtained by taking average spatial integrals over a period of periodic functions. 
If we define $ \oint \equiv (\prod L_i)^{-1}\int_{\{0\}}^{\{ {\bf L_i}\}}d{\bf x}$ then we have
\begin{align}\label{defgf2arg}
G_{AB}(\omega,\mathbf{k})\equiv G^{(\{0\})}_{AB}(\omega,\mathbf{k})=
\oint d\mathbf{x}\int d\mathbf{x}^{\prime}\,
G_{AB}(\omega,\mathbf{x},\mathbf{x}^{\prime})e^{i\mathbf{k}(\mathbf{x}^{\prime}-\mathbf{x})}\,.
\end{align}
From \eqref{eq:g_fourier} we can also write 
\begin{align}\label{defgf2arg2}
G_{AB}(\omega,\mathbf{k})
=(N\prod_i L_i)^{-1}G_{AB}(\omega,\mathbf{k},\mathbf{k})\,,
\end{align}
where $N$ is the total number of spatial periods in the system.

We next examine the positivity of the spectral weight of our operators. Working in the interaction picture,
the system absorbs energy at rate
\begin{align}\label{eq:abs_rate}
\frac{d}{dt}W(t)=\int d\mathbf{x}\,\delta\langle{B}\rangle (t,\mathbf{x})\,\frac{d}{dt}\delta h_{B}(t,\mathbf{x})\,,
\end{align}
where a summation over $B$ is understood. Introducing the notation
\begin{align}
\delta h_{B}(t,\mathbf{x})&=\frac{1}{(2\pi)^{d+1}}\int d\omega d\mathbf{k}\,\delta h_{B}(\omega,\mathbf{k})\,e^{-i\omega t+i\mathbf{k}\mathbf{x}}\,,
\end{align}
we can show that the total energy absorbed by the system is
\begin{align}
\Delta W
=&-\frac{1}{(2\pi)^{2d+1}}\int d\omega d\,\mathbf{k}d\mathbf{k}^{\prime}
\delta h^{\ast}_{B}(\omega,\mathbf{k})
\omega [\mathrm{Im}G]_{BB^{\prime}}(\omega,\mathbf{k},\mathbf{k}^{\prime})
\delta h_{B^{\prime}}(\omega,\mathbf{k}^{\prime})\,,
\end{align}
where $ [\mathrm{Im}G]_{AB}(\omega,\mathbf{k},\mathbf{k}^{\prime})\equiv 
\frac{1}{2i}[G_{AB}(\omega,\mathbf{k},\mathbf{k}^{\prime})
 - G^*_{BA}(\omega,\mathbf{k}^{\prime},\mathbf{k})]$. To get to the last line we used
 $G_{AB}(\omega,\mathbf{k},\mathbf{k}^{\prime})=G_{AB}(-\omega,-\mathbf{k},-\mathbf{k}^{\prime})^{\ast}$ (for real frequencies and wavevectors), which follows from the reality of $G_{AB} (t,\mathbf{x},\mathbf{x}^{\prime})$. 
 Since $\delta h_{B}(\omega,\mathbf{k})$ are arbitrary we deduce that 
 $-\omega[\mathrm{Im}G]_{AB}(\omega,\mathbf{k},\mathbf{k}^{\prime})$ is a positive semi-definite matrix, 
 with matrix indices including both the operator labels as well as the wavevectors.
Since the block diagonal elements of a positive semi-definite matrix are positive semi-definite,
 using \eqref{defgf2arg2} we can conclude that
the zero modes $-\omega Im G_{AB}(\omega,\mathbf{k})$ are positive semi-definite.
In particular we have
 \begin{align}\label{posprop}
 -\omega Im G_{AA}(\omega,\mathbf{k})\ge 0\,,
 \end{align}
 with no sum on $A$.
 The positive semi-definite aspect of 
 $-\omega[\mathrm{Im}G]_{AB}(\omega,\mathbf{k},\mathbf{k}^{\prime})$ also gives rise to additional conditions for the
 $G^{(\{n_{j}\})}_{AB}(\omega,\mathbf{k})$, with $\{n_{j}\}\ne \{0\}$.
 
To conclude this subsection we examine how the Green's functions behave under time reversal invariance. 
For simplicity we will assume that the periodic system is invariant under time reversal.
Recall that this acts
on local operators according to $T\,A(t,\mathbf{x})\,T^{-1}=\epsilon_{A} A(-t,\mathbf{x})$, where $\epsilon_A=\pm 1$. 
Since $T$ is an anti-unitary operator we can deduce that
$G_{AB}(t,\mathbf{x},\mathbf{x}^{\prime})=\epsilon_{A}\epsilon_{B}\,G_{BA}(t,\mathbf{x}^{\prime},\mathbf{x})$. Thus, we have
$G_{AB}(\omega,\mathbf{k},\mathbf{k}^{\prime})=\epsilon_{A}\epsilon_{B}\,G_{BA}(\omega,-\mathbf{k}^{\prime},-\mathbf{k})$
and hence
 \begin{align}\label{trevcon}
G^{(\{n_{j}\})}_{AB}(\omega,\mathbf{k})=\epsilon_{A}\,\epsilon_{B}\,G^{(\{n_{j}\})}_{BA}(\omega,-\mathbf{k}-n_{l}\mathbf{k}^{l}_{L})\,.
\end{align}

Returning to the linear response given in \eqref{eq:response}, after taking suitable Fourier transforms we can write
\begin{align}\label{eq:response2}
\delta\vev{{A}}(\omega,\mathbf{x})&=\frac{1}{(2\pi)^{2d}}\int d\mathbf{k}\,\sum_{\{n_{j}\}}
e^{i(\mathbf{k}+n_{j}\,\mathbf{k}_{L}^{j})\mathbf{x}}
G^{(\{n_{j}\})}_{AB}(\omega,\mathbf{k})
\delta h_{B}(\omega,\mathbf{k})\,.
\end{align}
If we consider a source which contains a single spatial Fourier mode
$\delta h_B(t,\mathbf{x})=e^{i\mathbf{k}_s\mathbf{x}}\delta h_B(t)$, then we have
\begin{align}\label{eq:response3}
\delta\vev{{A}}(\omega,\mathbf{x})&=
e^{i\mathbf{k}_s\mathbf{x}}
\sum_{\{n_{j}\}}\frac{1}{(2\pi)^d}
e^{in_{j}\,\mathbf{k}_{L}^{j}\mathbf{x}}
G^{(\{n_{j}\})}_{AB}(\omega,\mathbf{k}_s)
\delta h_{B}(\omega)\,,\nn
&\equiv e^{i\mathbf{k}_s\mathbf{x}}
\sum_{\{n_{j}\}}e^{i n_{j}\mathbf{k}_{L}^{j}\mathbf{x}}\, \delta\vev{A}^{(\{n_{j}\})}(\omega,\mathbf{k})\,.
\end{align}
Notice, in particular, that the zero mode in the summation is fixed by the zero mode of the Green's function:
$\delta\vev{A}^{(\{0\})}(\omega,\mathbf{k})=(2\pi)^{-d}G_{AB}(\omega,\mathbf{k}_s)
\delta h_{B}(\omega)$.

In the next sub-sections we will take $A$ and $B$ to be components of conserved currents. In this
context the zero-mode correlator $G_{AB}(\omega,\mathbf{k})$ captures transport of the associated hydrodynamic modes and
hence one can call it a `hydrodynamic-mode correlator'.

\subsection{Einstein relation for a single current}\label{secsc}

We now consider the operator $A$ to be a current density\footnote{In this section we find it convenient to work with current vector densities. In section \ref{relhsec} we will work with current vectors. We also note that as our analysis will focus on two-point functions of the current, we only
require that the current to be conserved at the linearised level.} 
operator $J^\mu$, which satisfies a continuity equation of the form $\partial_\mu J^\mu=0$. From the definition \eqref{eq:g_fourier} we have
\begin{align}\label{eq:fourier_ward}
-i\omega\, G_{J^{t}B}(\omega,\mathbf{k},\mathbf{k}^{\prime})+i\mathbf{k}_{i}G_{J^{i}B}(\omega,\mathbf{k},\mathbf{k}^{\prime})=0\,,
\end{align}
for any operator ${B}$, whose equal time commutator with $J^t$ vanishes. 
Using the crystallographic decomposition \eqref{eq:g_cystal} in \eqref{eq:fourier_ward}
we then have
\begin{align}\label{eq:ward_f_space}
-i\omega\, G^{(\{n_{j}\})}_{J^{t}B}(\omega,\mathbf{k})+i\left(\mathbf{k}+n_{j}\mathbf{k}^{j}_{L}\right)_{i}\,G^{(\{n_{j}\})}_{J^{i}B}(\omega,\mathbf{k})=0\,.
\end{align}

We now\footnote{We have also presented some more general results in appendix \ref{genrsa}.}
focus on the hydrodynamic-mode correlators with $\{n_{j}\}=0$, which satisfy a positivity property discussed just above \eqref{posprop}. Using \eqref{eq:ward_f_space} twice, we have
\begin{align}
-i\omega\, G_{J^{t}J^{t}}(\omega,\mathbf{k})+i\mathbf{k}_{i}G_{J^{i}J^{t}}(\omega,\mathbf{k})=&0\,,\notag\\
-i\omega\, G_{J^{t}J^{j}}(\omega,\mathbf{k})+i\mathbf{k}_{i}G_{J^{i}J^{j}}(\omega,\mathbf{k})=&0\label{eq:j_ide}\,.
\end{align}
We next consider the time reversal invariance conditions \eqref{trevcon} with $\{n_{j}\}=0$.
Since $\epsilon_{J^t}=+1$ and $\epsilon_{J^i}=-1$, we obtain
\begin{align}\label{eq:t_reversal}
G_{J^{i}J^{t}}(\omega,\mathbf{k})=-G_{J^{t}J^{i}}(\omega,-\mathbf{k})\,,\notag\\
G_{J^{i}J^{j}}(\omega,\mathbf{k})=G_{J^{j}J^{i}}(\omega,-\mathbf{k})\,.
\end{align}
Combing \eqref{eq:t_reversal} with \eqref{eq:j_ide} we therefore have the key result
\begin{align}\label{krone}
\frac{1}{i\omega} \mathbf{k}_{i}\mathbf{k}_{j}\,G_{J^{i}J^{j}}(\omega,\mathbf{k})=-i\omega\,G_{J^{t}J^{t}}(\omega,\mathbf{k})\,.
\end{align}

In general, taking the $\omega\to 0$ limit of the correlator $G_{AB}(\omega,\mathbf{k})$ gives rise to a static, thermodynamic susceptibility. It will be useful to write
\begin{align}\label{defcsus}
-\lim_{\omega\to 0+i0}G_{J^{t}J^{t}}(\omega,\mathbf{k})\equiv\chi(\mathbf{k})\,,
\end{align}
where $\chi(\mathbf{k})$ is a charge-charge susceptibility (the sign here is explained in appendix \ref{genrsa}). 
Note that \eqref{krone} implies
\begin{align}
\lim_{\omega \to 0+i0}\frac{1}{\omega^2} \mathbf{k}_{i}\mathbf{k}_{j}\,G_{J^{i}J^{j}}(\omega,\mathbf{k})=-\chi(\mathbf{k})\,,
\end{align}
and in particular, the longitudinal part of the current-current susceptibility vanishes, $\lim_{\omega\to 0+i0}\mathbf{k}_{i}\mathbf{k}_{j}G_{J^{i}J^{j}}(\omega,\mathbf{k})=0$, provided that $\chi(\mathbf{k})$ is finite\footnote{Note that for a superfluid one can have $\chi(\mathbf{k})$ diverging
at $\mathbf{k}\to 0$.}, which we will assume.

In order to focus on studying the response to long wavelength sources, it will be convenient
to now rescale the wave-number ${\bf k}$ by $\varepsilon$ and write \eqref{krone} in the form
\begin{align}\label{eq:main1}
\frac{1}{i\omega} \mathbf{k}_{i}\mathbf{k}_{j}\,G_{J^{i}J^{j}}(\omega,\varepsilon\,\mathbf{k})=-\frac{i\omega}{\varepsilon^{2}}\,G_{J^{t}J^{t}}(\omega,\varepsilon\mathbf{k})\,.
\end{align}
We next note that the AC conductivity matrix is defined by taking the following limit of the transport correlators
\begin{align}
\sigma^{ij}(\omega)=-\lim_{\varepsilon\to 0}\frac{1}{i\omega} G_{J^{i}J^{j}}(\omega,\varepsilon\,\mathbf{k})\,.
\end{align}
Notice from the discussion above \eqref{posprop} that the real part of $\sigma^{ij}(\omega)$ is a positive semi-definite matrix.
In general, the AC conductivity is a finite quantity for $\omega\ne 0$. On the other hand, the DC conductivity, defined by $\sigma^{ij}_{DC}\equiv \lim_{\omega\to 0} \sigma(\omega)$, is not necessarily finite. For example, if the system is translationally invariant or
if the  breaking of translation invariance has arisen spontaneously, 
or more generally if there are Goldstone modes present, generically the DC conductivity will be infinite, or more precisely there will be a delta function on the AC conductivity at $\omega=0$. 
By taking the limit $\varepsilon\to 0$ in \eqref{eq:main1} we have
\begin{align}\label{eq:Gtt_restr}
\mathbf{k}_{i}\mathbf{k}_{j}\,\sigma^{ij}(\omega)&=i\omega\,\lim_{\varepsilon\to 0}\frac{1}{\varepsilon^{2}}\,G_{J^{t}J^{t}}(\omega,\varepsilon\mathbf{k})\,.
\end{align}
Thus when the $DC$ conductivity is finite, the function $G_{J^{t}J^{t}}(\omega,\varepsilon\mathbf{k})/\varepsilon^{2}$ must have a pole at $\omega=0$ after taking the $\varepsilon\to 0$ limit. Note that \eqref{defcsus} shows that 
before the limit is taken this pole is absent (provided that $\chi(\mathbf{k})$ is finite).

To make further progress, it is helpful to write
\begin{align}\label{geenen}
{G}_{J^tJ^t}(\omega,\varepsilon\mathbf{k})\chi(\varepsilon\mathbf{k})^{-1}=\frac{-N(\omega,\varepsilon\mathbf{k})}{-i\omega+N(\omega,\varepsilon\mathbf{k})}\,,
\end{align}
where we have defined the quantity
\begin{align}\label{enndef}
N(\omega,\varepsilon\mathbf{k})=\frac{{G}_{J^tJ^t}(\omega,\varepsilon\mathbf{k})}{\frac{1}{i\omega}({G}_{J^tJ^t}(\omega,\varepsilon\mathbf{k})+\chi(\varepsilon\mathbf{k}))}\,.
\end{align}
We can now prove that $N(\omega,\varepsilon\mathbf{k})$ is an analytic function of $\omega$ provided that $Im(\omega)\ne 0$. Firstly, 
any poles in the numerator ${G}_{J^tJ^t}(\omega,\varepsilon\mathbf{k})$, which can only occur in the lower half plane, will cancel out with those in the denominator. We thus need to check whether or not the denominator in \eqref{enndef} can vanish for $Im(\omega)\ne 0$. That this cannot occur can be seen by writing
\begin{align}
\frac{1}{i\omega}({G}_{J^tJ^t}(\omega,\varepsilon\mathbf{k})+\chi(\varepsilon\mathbf{k}))
=\int_{C_1}\frac{d\omega'}{i\pi}\frac{Im{G}_{J^tJ^t}(\omega',\varepsilon\mathbf{k})}{\omega'(\omega'-\omega)}\,,
\end{align}
where $C_1$ is a contour that skirts just under the real axis.
Then writing $\omega =x+i y$, with $y\ne 0$, we can show that the real part of the integral is non-vanishing
after using the fact that $Im{G}_{J^tJ^t}(\omega',\varepsilon\mathbf{k})/\omega'\le 0$, which we showed in \eqref{posprop}.
We now return to \eqref{eq:Gtt_restr}, from which we deduce that, for fixed $\omega$, as $\varepsilon\to 0$, we can expand 
\begin{align}\label{ennexp}
N=\varepsilon^2\frac{\mathbf{k}_{i}\mathbf{k}_{j}\,\sigma^{ij}(\omega)}{\chi{(\bf 0})}+\dots \,,
\end{align}
with the neglected terms going to zero with a higher power of $\varepsilon$.

We are now in a position to discuss the poles of ${G}_{J^tJ^t}(\omega,\varepsilon\mathbf{k})$ that appear at the `origin', by which we
mean when both $\omega\to 0$ and $\varepsilon\to 0$. The simplest possibility is if $N(\omega,\varepsilon\mathbf{k})$ does not have any
poles (or branch cuts) at $\omega =0$. In this case, we see that when the DC conductivity matrix is finite,
${G}_{J^tJ^t}$ will have a single diffusion pole
with dispersion relation
\begin{align}\label{feinr}
\omega=-i\varepsilon^2D(k)+\dots\,, \qquad D(k)=[\sigma^{ij}_{DC}k_ik_j]\chi({\bf 0})^{-1}\,,
\end{align}
and the neglected terms are higher order in $\varepsilon$.
This is our first result on Einstein relations for inhomogeneous media.

It is important to emphasise that is not the only possibility. Indeed, as we discuss in the next subsection, there are additional
poles when there are additional conserved currents.
If, for example, we suppose that there are two conserved currents in total then a second diffusion pole
can appear in ${G}_{J^tJ^t}(\omega,\varepsilon\mathbf{k})$. To illustrate this situation schematically, consider the behaviour of the following 
function for $\omega,\varepsilon {\bf k}\to 0$,
\begin{align}
\varepsilon^2\left(\frac{A}{-i\omega+\varepsilon^2 a}  
+\frac{B}{-i\omega+\varepsilon^2 b}  \right)
\sim\frac{\varepsilon^2(A+B)}{-i\omega+\varepsilon^2\left(\frac{aA+bB}{A+B}-i\frac{AB(a-b)^2}{(A+B)^2}\frac{\varepsilon^2}{\omega}+\mathcal{O}(\frac{\varepsilon^2}{\omega})^2\right)}\,.
\end{align}
corresponding to the function $N(\omega,\varepsilon\mathbf{k})$ having additional singularities at $\omega\to 0$. Another interesting
situation in which additional poles will appear is in the presence of Goldstone modes arising from broken symmetries. 
Additional general statements can be made using the memory matrix formalism, generalising the discussion in \cite{Forster}.

Returning now to the case in which there is just a single conserved current with a single diffusion pole 
then a natural phenomenological expression for the Green's function is given near the origin, $(\omega,\varepsilon{\bf k})\to 0$, by
\begin{align}\label{pheno}
G_{J^{t}J^{t}}(\omega,\varepsilon\mathbf{k})&\sim\frac{-D(\omega,\varepsilon\mathbf{k})}{-i\omega+D(\omega,\varepsilon\mathbf{k})}\chi(\varepsilon\mathbf{k})\,,
\end{align}
with $D(\omega,\varepsilon\mathbf{k})\sim\varepsilon^{2}\frac{\mathbf{k}_{i}\mathbf{k}_{j}\,\sigma^{ij}(\omega)}{\chi(\varepsilon\mathbf{k})}$. 
It is interesting to note that if, by contrast, we are in the context of infinite DC conductivity with 
$\sigma^{ij}(\omega)\sim K^{ij}\,\left(\frac{i}{\omega}+\pi\,\delta(\omega) \right)$ for small $\omega$, where $K^{ij}$ is constant,
then \eqref{pheno} is gives rise to sound modes for the current 
density $J^{t}$, with dispersion relation $\omega_{\pm}=\pm\varepsilon \sqrt{K^{ij}\,\mathbf{k}_{i}\mathbf{k}_{j}/\chi(\varepsilon\mathbf{k})}$. 
The transition between diffusion modes and sound modes was also discussed in a homogeneous hydrodynamic setting, with a phenomenological
term to relax momentum, in \cite{Davison:2014lua}.

To conclude this subsection, we briefly note that we can carry out a similar analysis for the higher Fourier modes of the current-current correlators.
Starting with \eqref{eq:ward_f_space}, the analogue of \eqref{eq:main1} is
\begin{align}\label{krone2}
\frac{1}{i\omega} (\mathbf{k}+n_{r}\mathbf{k}^{r}_{L})_{i}\,\mathbf{k}_{j}\,G^{(\{n_{l}\})}_{J_{A}^{i}J_{B}^{j}}(\omega,\mathbf{k})&=-i\omega\,G^{(\{n_{l}\})}_{J_{A}^{t}J_{B}^{t}}(\omega,\mathbf{k})\,,
\end{align}
and this leads, {\it mutatis-mutandis}, to additional relations concerning the poles of $G^{(\{n_{l}\})}_{J_{A}^{t}J_{B}^{t}}(\omega,\mathbf{k})$, which would
be interesting to explore in more detail. We note however, that for ${\{n_{l}\}}\ne 0$, there is no longer a simple statement concerning the
positivity of $Im G^{(\{n_{l}\})}_{J_{A}^{t}J_{B}^{t}}(\omega,\mathbf{k})/\omega$, which was used in the above. We also point out that
within a holographic context and for a specific gravitational model, some of the $G^{(\{n_{l}\})}_{J_{A}^{i}J_{B}^{j}}(\omega,\mathbf{k})$
were calculated in 
\cite{Donos:2014yya}.

\subsection{Generalised Einstein relations for multiple currents}

We now assume that we have multiple conserved currents $J^{\mu}_{A}$. For example, one could have both a conserved heat current
and a conserved $U(1)$ current.
Much of the analysis that we carried out for the case of a single current goes through straightforwardly and we obtain
\begin{align}\label{eq:main2}
\frac{1}{i\omega} \mathbf{k}_{i}\mathbf{k}_{j}\,G_{J_{A}^{i}J_{B}^{j}}(\omega,\varepsilon\,\mathbf{k})=-\frac{i\omega}{\varepsilon^{2}}\,G_{J_{A}^{t}J_{B}^{t}}(\omega,\varepsilon\mathbf{k})\,.
\end{align}
We write the charge susceptibilities and the AC conductivity via 
\begin{align}\label{eq:Gtt_restr2}
\boldsymbol\chi_{AB}(\varepsilon\mathbf{k})&=-\lim_{\omega\to 0+i0}G_{J_{A}^{t}J_{B}^{t}}(\omega,\varepsilon\mathbf{k})\,,\nn
\sigma^{ij}_{AB}(\omega)&=-\lim_{\varepsilon\to 0}\frac{1}{i\omega} G_{J^{i}_AJ^{j}_B}(\omega,\varepsilon\,\mathbf{k})\,,
\end{align}
respectively, and we now have 
\begin{align}\label{condsecond}
\mathbf{k}_{i}\mathbf{k}_{j}\,\sigma_{AB}^{ij}(\omega)&=i\omega\,\lim_{\varepsilon\to 0}\frac{1}{\varepsilon^{2}}\,G_{J_{A}^{t}J_{B}^{t}}(\omega,\varepsilon\mathbf{k})\,.
\end{align}
Generically this shows that for finite DC conductivities there will be at least as many poles in the transport current correlators
as there are currents.

Proceeding much as before we write
\begin{align}
\mathbf{G}(\omega,\varepsilon\mathbf{k})\boldsymbol\chi(\varepsilon\mathbf{k})^{-1}=-\left[-i\omega +{\bf N}(\omega,\varepsilon\mathbf{k})\right]^{-1}{\bf N}(\omega,\varepsilon\mathbf{k})\,,
\end{align}
where $\mathbf{G}(\omega,\varepsilon\mathbf{k})_{AB}\equiv G_{J_{A}^{t}J_{B}^{t}}(\omega,\varepsilon\mathbf{k})$ and
\begin{align}\label{enndef2}
{\bf N}(\omega,\varepsilon\mathbf{k})\equiv\mathbf{G}(\omega,\varepsilon\mathbf{k})\left[  \frac{1}{i\omega}(\mathbf{G}(\omega,\varepsilon\mathbf{k})
+\boldsymbol\chi(\varepsilon\mathbf{k}))   \right]^{-1}\,.
\end{align}
We can again argue that ${\bf N}(\omega,\varepsilon\mathbf{k})$ can only have poles on the real $\omega$ axis.
From \eqref{condsecond} we deduce that for fixed $\omega$, as $\varepsilon\to 0$, we can expand
\begin{align}
{\bf N}(\omega,\varepsilon\mathbf{k})=\varepsilon^2\mathbf{\Sigma}(\omega,\mathbf{k})\boldsymbol\chi(\varepsilon\mathbf{k})^{-1}\,,
\end{align}
where $\mathbf{\Sigma}(\omega,\mathbf{k})_{AB}=\mathbf{k}_{i}\mathbf{k}_{j}\,\sigma_{AB}^{ij}(\omega)$
and the neglected terms go to zero with a higher power of $\varepsilon$.

If we now assume that ${\bf N}(\omega,\varepsilon\mathbf{k})$ doesn't have any poles at $\omega=0$, then we can conclude
that at the origin, i.e. when both $\omega\to 0$ and $\varepsilon\to 0$, if the DC conductivities are finite
then the diffusion poles of the system are located at 
\begin{align}\label{geneinrel}
\omega_{A}(\mathbf{k})=-i\,D_{A}(\mathbf{k})\,\varepsilon^{2}+\cdots\,,
\end{align}
where $D_{A}(\mathbf{k})$ are the eigenvalues of what can be called the `generalised diffusion matrix' $\mathbf{D}(\mathbf{k})$ defined by
\begin{align}\label{geneinrel2}
\mathbf{D}(\mathbf{k})=\mathbf{\Sigma}(0,\mathbf{k})\boldsymbol\chi({\bf 0})^{-1}\,,
\end{align}
and the dots involve higher order corrections in $\varepsilon$.
In particular when the DC conductivities are finite, the number of diffusion poles is the same as the number of conserved currents.

Furthermore, we emphasise that when there is more than one conserved current, generically, these diffusion modes
do not satisfy a dispersion relation of the form $\omega\sim -i\varepsilon^2 \Sigma_{ij}k^ik^j$, with the matrix $\Sigma_{ij}$ a component of the DC conductivities. As a consequence we refer to our result \eqref{geneinrel},\eqref{geneinrel2} as a `generalised Einstein relation'.

We conclude this section by noting that the general result \eqref{geneinrel}, \eqref{geneinrel2} relates thermodynamic instabilities to dynamic instabilities. Suppose that the system has a static susceptibility matrix $\boldsymbol\chi({\bf 0})$ with
a negative eigenvalue and hence is thermodynamically unstable. Then \eqref{geneinrel2} 
implies that $\mathbf{D}(\mathbf{k})$ will have a
negative eigenvalue, for small $\mathbf{k}$, and hence, from \eqref{geneinrel} we deduce that there will be a diffusion pole
in the upper half plane leading to a dynamical instability\footnote{An explicit example of such a dynamic instability can be seen using the results of
appendix \ref{appa}.}.

\section{Diffusion in relativistic hydrodynamics}\label{relhsec}

We now discuss thermoelectric transport within the context of relativistic hydrodynamics. As well as generalising
the work of \cite{Banks:2016wdh} to include a conserved $U(1)$ charge (as also studied in \cite{Scopelliti:2017sga}),
we will also be able to use the formalism
to illustrate the results of the previous section. In particular, associated with the heat current and the $U(1)$ current we construct two
diffusion modes with dispersion relations satisfying the generalised Einstein relation \eqref{geneinrel}. 
We note that it will be convenient to use a slightly different notation
in this section, which implies that a little care is required in directly comparing with the last section.

\subsection{General setup}
We will consider an arbitrary relativistic quantum field theory with a global $U(1)$ symmetry in $d\ge 2$ spacetime dimensions.
The field theory is defined on a static, curved manifold, with metric $g_{\mu\nu}$, and a non-zero background gauge-field, $A_\mu$, 
of the form:
\begin{align}\label{bfield}
ds^2&=-f^2(x)dt^2+h_{ij}(x)dx^i dx^j\,,\nn
A_t&=a_t(x)\,.\end{align}
This corresponds to studying the field theory with $f^2$ and $h_{ij}$ parametrising 
sources for the stress tensor components $T^{tt}$ and $T^{ij}$, respectively, and 
$a_t$ parametrising a source for the $J^t$ component of the conserved $U(1)$ current.
We focus on cases in which the manifold has planar topology, with the globally defined spatial coordinates $x^i$ parametrising $\mathbb{R}^{d-1}$,
and $f$, $h_{ij}$,$a_t$ all 
depending periodically on $x^i$, with period $L_i$.

We will study the field theory at finite temperature in the hydrodynamic limit keeping the leading order viscous terms. 
In particular, we will consider temperatures\footnote{This temperature is the same as what is denoted as $\bar T_0$ below.}
that are much greater than the largest wave-number that appears in the background fields in \eqref{bfield}.
The Ward identities are given by
\begin{align}\label{wids}
D_\mu T^{\mu\nu}=F^{\nu\lambda}J_{\lambda}\,,\qquad D_{\mu} J^{\mu}=0\,, 
\end{align}
where $D_\mu$ is the covariant derivative with respect to $g_{\mu\nu}$ and
$F_{\mu\nu}=2\partial_{[\mu}A_{\nu]}$.  For the special case of conformal field theory, we should also impose $T^\mu{}_\mu=0$ and this
implies, amongst other things, that in \eqref{eq:EMTe} $\zeta_b=0$ and $\epsilon=(d-1)P$.

The hydrodynamic variables are the local temperature, $T(x)$, the local chemical potential, $\mu(x)$, and the fluid velocity, $u^\mu$,
with $u^\mu u^\nu g_{\mu\nu}=-1$.
As in \cite{Jensen:2011xb}, the constitutive relations are given, in the Landau frame,
 by\footnote{Following \cite{Jensen:2011xb}, we have set to 
zero two other terms in $J^\mu$ that are allowed by Lorentz invariance but are not consistent with positivity of entropy and
thermodynamics with external sources.}
\begin{align}
\label{eq:EMTe}
T_{\mu \nu} = &P g_{\mu \nu} + (P+\epsilon)u_{\mu}u_{\nu} - 2 \eta\left( D_{(\mu} u_{\nu)} + u_{\rho}u_{(\mu} D^{\rho} u_{\nu)} - (g_{\mu \nu} +u_{\mu}u_{\nu}) \frac{D_{\rho} u^{\rho}}{d-1}\right)\nn
& -\zeta_{b}(g_{\mu\nu}+u_{\mu}u_{\nu})D_{\rho}u^{\rho}\,,\nn
J^{\mu}  =& \rho u^{\mu} + \sigma_Q \left(F^{\mu\nu}u_{\nu} - T(g^{\mu \nu} + u^{\mu}u^{\nu})D_{\nu} \left(\frac{\mu}{T}\right)\right) \,,
\end{align}
where $P$ is the pressure density, $\epsilon$ is the energy density and $\rho$ is the $U(1)$ charge density.
The dissipative terms in \eqref{eq:EMTe} are the shear viscosity, $\eta$, the bulk viscosity, $\zeta_{b}$ and the conductivity, $\sigma_Q$, which
should not be confused with the electrical DC conductivity, $\sigma_{DC}$, which we discuss later.
We also have the local thermodynamic relation and first law which take the form
\begin{align}\label{firstlaw}
P+\epsilon=sT+\mu \rho,\qquad d P=sd T+\rho d\mu\,,
\end{align}
where $s$ is the entropy density.
It will also be helpful to introduce the susceptibilities $c_{\mu}$, $\xi$ and $\chi$ via
\begin{align}\label{eq:suscepdef1}
ds=T^{-1}c_{\mu}\,d T+\xi\,d\mu\,,\qquad
d \rho=\xi\,d T+\chi\,d\mu \,.
\end{align}

For any vector $k$, the Ward identities imply 
\begin{align}
D_\mu[(T^\mu{}_\nu+J^\mu A_\nu)k^\nu]=\frac{1}{2}\mathcal{L}_kg_{\mu\nu}T^{\mu\nu}+\mathcal{L}_kA_\mu J^\mu\,,
\end{align}
where $\mathcal{L}_k$ is the Lie derivative. 
Taking $k=\partial_t$ we define the heat current as 
\begin{align}\label{qexp}
Q^\mu=-(T^\mu{}_t+A_tJ^\mu)\,,
\end{align}
which is conserved for stationary metrics with $\mathcal{L}_kA_\nu=0$. Thus, given such background metrics and gauge fields,
for time independent configurations we therefore have $\partial_i(\sqrt{-g}\, Q^i)=\partial_i (\sqrt{-g}\,J^i)=0$.

In thermal equilibrium the fluid configuration is given by
\begin{align}
u_t&=-f(x)\,,\qquad u_i=0\,,\qquad T=T_0(x)\,,\qquad \mu=\mu_0(x)\,,
\end{align}
where $T_0(x)$ and $\mu_0(x)$ are periodic functions,
and from \eqref{firstlaw} we have the equilibrium relations
\begin{align}\label{flaweq}
P_0+\epsilon_0=s_0T_0+\mu_0 \rho_0,\qquad \partial_i P_0=s_0\partial_i T_0+\rho_0\partial_i\mu_0\,.
\end{align}
For later use, we note that we also have
\begin{align}\label{eq:suscepdef2}
\nabla_{i}s_{0}=T^{-1}_{0}c_{\mu0}\,\nabla_{i}T_{0}+\xi_0\,\nabla_{i}\mu_{0}\,,\qquad
\nabla_{i}\rho_{0}=\xi_0\nabla_{i}T_{0}+\chi_0\,\nabla_{i}\mu_{0}\,.
\end{align}
By calculating $T^{\mu\nu}$, $J^{\mu}$ one can show that the Ward identities are satisfied provided that
\begin{align}\label{fTatmu}
T_{0}=f^{-1}\bar{T}_{0}\,,\qquad
\mu_{0}= f^{-1}a_{t}\,,
\end{align}
where $\bar{T}_{0}$ is constant. Note, in particular, that in thermal equilibrium the local hydrodynamic variable $T_0$ is
not constant when $f$ is not constant and, furthermore, there is a factor of $f$ that appears in the relationship
between $\mu_0$ and the background gauge field. We also note that we have set a possible integration constant to zero
in the second expression as we want $\mu_0$ to vanish when $a_t$ does. Finally it will be helpful to define the zero mode of $a_t$
via $\bar\mu_0\equiv \oint a_t$,  
where we are again using the notation
 $\oint \equiv (L_1\cdots L_d)^{-1} \int_{\{0\}}^{\{ {\bf L_i}\}} dx^1 \cdots dx^d$.
This allows us to write $\mu_0=f^{-1}(\bar\mu_0+\tilde a_t(x))$, with $\oint \tilde a_t=0$.

The non-vanishing components of the stress tensor and current for this equilibrium configuration are then given by
\begin{align}\label{tjperturbativeeq}
T_{tt}=\epsilon_{0}f^{2},\qquad
T_{ij}=P_{0}h_{ij},\qquad
J^{t}=\rho_{0}f^{-1}\,.
\end{align}
In particular for the backgrounds we are considering, in thermal equilibrium both the electric and the
heat currents vanish: $J^i=Q^i=0$. Note, since \eqref{bfield} provides a source for the energy and the charge, we can immediately
deduce that the charge-current susceptibilities must vanish. The total energy and charge of the equilibrium configuration
are defined by 
\begin{align}
\epsilon_{tot}&=-\oint\sqrt{-g}T^{t}{}_t=\oint\sqrt{h}f\epsilon_0\,,\nn
\rho_{tot}&=\oint\sqrt{-g}J^t=\oint\sqrt{h}\rho_0\,.
\end{align}
We can also define the total equilibrium entropy as
\begin{align}
s_{tot}=\oint\sqrt{h} s_0\,.
\end{align}
For later use, using the fact that $s_0$ is a function of $T_0$ and $\mu_0$, 
we observe that for suitable zero modes of the charge susceptibilities we have
\begin{align}\label{totsus}
\frac{\partial s_{tot}}{\partial \bar T_0}
= \oint \sqrt{h} f^{-1}T_0^{-1}c_{\mu0}\,,\qquad
\frac{\partial s_{tot}}{\partial \bar \mu_0}
= \oint \sqrt{h} f^{-1}\xi_0\,.
\end{align}
Similarly, we also have
\begin{align}\label{totsus2}
\frac{\partial \rho_{tot}}{\partial \bar T_0}
= \oint \sqrt{h} f^{-1}\xi_{0}\,,\qquad
\frac{\partial \rho_{tot}}{\partial \bar \mu_0}
= \oint \sqrt{h} f^{-1}\chi_0\,.
\end{align}

\subsection{Generalised Navier-Stokes equations}
In the following we want to study the behaviour of small perturbations about the equilibrium configuration, including the possibility of
adding external, perturbative thermal gradient and electric field sources. Following \cite{Banks:2016wdh} we will do this by considering
\begin{align}\label{pert}
ds^{2}&=-f^{2}(1-2\phi_{T})\,dt^{2}+h_{ij}dx^{i}dx^{j}\,,\nn
A_{t}&=a_{t}-f\mu_{0}\phi_{T}+\phi_{E}\,,
\end{align}
along with
\begin{align}
u_t&=-f(1-\phi_{T})\,,\qquad u_i=\delta u_i\,,\nn
T&=T_0+\delta T\,,\qquad\qquad \mu=\mu_0+\delta\mu\,.
\end{align}
Here $\phi_T$, $\phi_E$, $\delta u_i$, $\delta T$ and $\delta\mu$ are all functions
of $(t,x^i)$.
Note that these need not be periodic functions of the spatial coordinates.
For later use, we also define the spatial components of the external sources $\zeta_i$, $E_i$ via
\begin{align}
\zeta_i=\partial_i \phi_T,\qquad E_i=\partial_i \phi_E\,.
\end{align}

At linearised order, the perturbed stress tensor and $U(1)$ current can then be written as
\begin{align}\label{tjperturbative}
T_{tt}&=\epsilon_{0}f^{2}(1-2\phi_{T})+\delta\epsilon\,f^{2}\,,\notag\\
T_{ti}&=-f(P_{0}+\epsilon_{0})\delta u_{i}\,,\notag\\
T_{ij}&=(P_{0}+\delta P)h_{ij}-2\eta_{0}f^{-1}\left(\nabla_{(i}\left(f\delta u_{j)}\right)-\frac{h_{ij}}{(d-1)}\nabla_{k}\left(f\delta u^{k}\right)\right)-\zeta_{b0}h_{ij}f^{-1}\nabla_{k}\left(f\delta u^{k}\right)\,,\notag\\
J^{t}&=\rho_{0}f^{-1}(1+\phi_{T})+f^{-1}\delta \rho\,,\notag\\
J^{i}&=\rho_{0}\delta u^{i} + \sigma_{Q0}f^{-1}\left[E^{i}-\nabla^{i}\left(f\delta\mu\right) -f\mu_{0}\zeta^{i}+\mu_{0}T_{0}^{-1}\nabla^{i}\left(f\delta T\right)\right]\,,
\end{align}
where $\nabla_i$ is the covariant derivative with respect to the metric $h_{ij}$, which is also used to raise and lower indices.
The Ward identities \eqref{wids} give
\begin{align}\label{thermoelectricstokesgeneralo1}
&\partial_{t}\delta \rho + \nabla_{i}\left(fJ^{i}\right)=0\,,\notag\\
&f\partial_t\delta\epsilon + \nabla_{i}\left(f^{2}(P_{0}+\epsilon_{0})\delta u^{i}\right)-fJ^{i}\nabla_{i}a_{t}=0\,,\notag\\
&f^{-1}(P_{0}+\epsilon_{0})\partial_{t}\delta u_{j} -2f^{-1}\nabla^{i}\left(\eta_{0}\nabla_{(i}\left(f\delta u_{j)}\right)\right) + f^{-1}\nabla_{j}\left(\left(\frac{2\eta_{0}}{(d-1)}-\zeta_{b0}\right)\nabla_{k}\left(f\delta u^{k}\right)\right)\notag\\
&=-\nabla_{j}\delta P -(\delta\epsilon+\delta P)f^{-1}\nabla_{j}f+(P_{0}+\epsilon_{0})\zeta_{j} + \rho_{0}(f^{-1}E_{j}-\mu_{0}\zeta_{j})+f^{-1}\delta \rho\nabla_{j}a_{t}\,,
\end{align}
In the case when there is no $U(1)$ charge this agrees with the expression derived in equation (A.10) of \cite{Banks:2016wdh}.
These expressions can be further simplified. We use \eqref{fTatmu} as well as
\begin{align}
\delta P=s_0\delta T+\rho_0\delta \mu\,,\qquad \delta\epsilon=T_0\delta s+\mu_0\delta\rho\,,\nn
\delta s=T^{-1}_{0}c_{\mu0}\,\delta T+\xi_0\,\delta\mu,\qquad
\delta \rho=\xi_0\,\delta T+\chi_0\,\delta\mu \,,
\end{align}
which we obtain from \eqref{firstlaw},\eqref{eq:suscepdef1}. After also using \eqref{eq:suscepdef2} we eventually find that
we can rewrite the system \eqref{thermoelectricstokesgeneralo1} in the following form, which is the key result
of this section,
\begin{align}\label{testokes}
\xi_0\partial_{t}\delta T + \chi_0\partial_{t}\delta\mu + \nabla_{i}\left(fJ^{i}\right)&=0\,,\notag\\
fc_{\mu0}\partial_{t}\delta T + fT_{0}\xi_0\partial_{t}\delta\mu + \nabla_{i}\left(fQ^{i}\right)&=0\,,\notag\\
(P_{0}+\epsilon_{0})\partial_{t}\delta u_{j} -2\nabla^{i}\left(\eta_{0}\nabla_{(i}\left(f\delta u_{j)}\right)\right) + \nabla_{j}\left(\left(\frac{2\eta_{0}}{(d-1)}-\zeta_{b0}\right)\nabla_{k}\left(f\delta u^{k}\right)\right)&=\notag\\
\rho_{0}\left[E_{j}-\nabla_{j}\left(f\delta\mu\right)\right] + fT_{0}s_{0}\left[\zeta_{j}-(fT_{0})^{-1}\nabla_{j}\left(f\delta T\right)\right]&\,,
\end{align}
with 
\begin{align}\label{constlin}
J^{i}&=\rho_{0}\delta u^{i} + \sigma_{Q0}f^{-1}\left[E^{i}-\nabla^{i}\left(f\delta\mu\right)\right] -\sigma_{Q0}\mu_{0}\left[\zeta^{i}-(fT_{0})^{-1}\nabla^{i}\left(f\delta T\right)\right]\,,\nn
Q^i&=f(P_{0}+\epsilon_{0})\delta u^{i}-f\mu_{0}J^{i}\,.
\end{align}
Notice that the first two lines in \eqref{testokes} are just current conservation equations for the linearised perturbation.
We emphasise that all background equilibrium quantities, marked with a $0$ subscript, 
are all periodic functions of the spatial coordinates.
It is interesting to note that the system of equations
\eqref{testokes} is invariant under the interchange
\begin{align}\label{spsymmetry}
E_{j} \leftrightarrow -\nabla_{j}\left(f\delta\mu\right)\,,\qquad
\zeta_{j} \leftrightarrow  -f^{-1}T_{0}^{-1}\nabla_{j}\left(f\delta T\right)\,.
\end{align}

Finally, for later use, we note that when the sources are set to zero, $\phi_T=\phi_E=0$, we have for the total
charges
\begin{align}\label{tchges}
\oint \sqrt{-g}J^t&=\oint\sqrt{h}\rho_0+ \oint\sqrt{h}\delta\rho\,,\nn
\oint \sqrt{-g}Q^t&=\oint\sqrt{h}f(\epsilon_0-\mu_0\rho_0)+\bar T_0\oint\sqrt{h}\delta s\,.
\end{align}

\subsection{Thermoelectric DC conductivity}

We now explain how we can obtain the thermoelectric DC conductivity, generalising \cite{Banks:2016wdh}.
We begin by considering the sources $\phi_T$ and $\phi_E$ to have space and time dependence 
of the form $e^{-i\omega t}e^{ik_ix^i}$, where $k_i$ is an arbitrary wave number. After solving 
\eqref{testokes} for $\delta u_{j}$, $\delta\mu$, $\delta T$ one obtains the local currents $J^i$ $Q^i$, and hence the current fluxes
$\bar J^i$ $\bar Q^i$, as functions of
$E_i$ and $\zeta_i$.  To obtain the thermoelectric DC conductivity we should then take the limit $k_i\to 0$, followed by $\omega\to 0$.

By considering approximating $e^{ik_ix^i}\sim 1+ik_ix^i$ 
we are prompted\footnote{An alternative procedure is to consider sources that are linear in time, as explained in a holographic context in
\cite{Donos:2014uba,Donos:2014cya}.}
 to consider a time-independent
source of the form
\begin{align}
\phi_T=x^i\bar \zeta_i,\qquad \phi_E=x^i\bar E_i\,,
\end{align}
where $\bar \zeta_i$, $\bar E_i$ are constants
and hence $E_i=\bar E_i$, $\zeta_i=\bar\zeta_i$. 
After substituting into \eqref{eq:hydro_fluid_exp2}
we obtain the system\footnote{In the special case of conformal field theories, similar equations were obtained in a holographic context in
\cite{Donos:2015gia}. The equations differ when there is a $U(1)$ symmetry due to a difference in the expression for
$Q^i$ in \eqref{constlin}. The equations should agree in the hydrodynamic limit, after a possible change of frame and/or incorporating higher order terms in the hydrodynamic expansion, and it would be interesting to investigate this in more detail.}
\begin{align}\label{testokeso0}
\nabla_{i}\left(fJ^{i}\right)=0\,,\qquad
\nabla_{i}\left(fQ ^{i}\right)=0\,,\notag\\
-2\nabla^{i}\left(\eta_{0}\nabla_{(i}\left(f\delta u_{j)}\right)\right) + \nabla_{j}\left(\left(\frac{2\eta_{0}}{(d-1)}-\zeta_{b0}\right)\nabla_{k}\left(f\delta u^{k}\right)\right)&=\notag\\
\rho_{0}\bar E_{j}
-\rho_{0}\nabla_{j}\left(f\delta\mu\right) + fs_{0}T_{0}\bar \zeta_{j}
&-s_{0}\nabla_{j}\left(f\delta T\right)&\,.
\end{align}
After solving these equations we obtain the local time-independent, steady state currents $J^i(x)$ $Q^i(x)$,
periodic in the spatial coordinate, as functions
of $\bar \zeta_i$, $\bar E_i$. We can now define the heat and charge current fluxes via
\begin{align}
\bar Q^i\equiv \oint \sqrt{-g}Q^i=\oint\sqrt{h} fQ^i\,,\qquad \bar J^i\equiv \oint\sqrt{-g} J^i=\oint\sqrt{h} fJ^i\,,
\end{align}
and the DC conductivities are obtained from
\begin{align}\label{dcmat}
\left( \begin{array}{c}
\bar J^i\\
\bar Q^i 
\end{array} \right)=
\left( \begin{array}{cc}
\sigma_{DC}^{ij} &\bar{T}_{0}\alpha_{DC}^{ij}\\
\bar{T}_{0}\bar{\alpha}_{DC}^{ij} &\bar{T}_{0}\bar{\kappa}_{DC}^{ij} 
\end{array} \right)
\left( \begin{array}{c}
\bar E_{j}\\
\bar\zeta_{j} 
\end{array} \right)\,.
\end{align}
Since we are considering backgrounds which preserve time reversal invariance the Onsager relations imply
that $\sigma_{DC}$ and $\bar\kappa_{DC}$ are symmetric matrices and $\alpha^T_{DC}=\bar\alpha_{DC}$.

\subsection{Diffusive modes}\label{diffmodes}
We now discuss how we can construct a perturbative diffusive solution of the system of equations \eqref{testokes}
that is associated with diffusion modes. Our objective will be to extract the associated dispersion relations for these modes.

We first set the source terms in \eqref{testokes} to zero: $E_i=\zeta_i=0$. We will allow for
a time-dependence of the form $e^{-i\omega t}$ and consider the expansion
\begin{align}\label{eq:hydro_fluid_exp}
\omega&= \sum_{\alpha=1}^{\infty} \varepsilon^\alpha \omega^{(\alpha)}\,,
\end{align}
with $\varepsilon\ll1 $. Since we are interested 
in wavelengths that are much larger than the periods, $L_i$, of the background fields in \eqref{bfield},
we introduce arbitrary wave numbers $k^i$ and consider
\begin{align}\label{eq:hydro_fluid_exp2}
\delta T&=e^{-i\omega t}e^{i \varepsilon k_i x^i} \sum_{\alpha=0}^{\infty} \varepsilon^\alpha\delta T^{(\alpha)}(x)\,, &&\delta \mu=e^{-i\omega t}e^{i \varepsilon k_i x^i} \sum_{\alpha=0}^{\infty} \varepsilon^\alpha \delta \mu^{(\alpha)}(x)\,,\notag\\
\delta u_i&=e^{-i\omega t}e^{i \varepsilon k_i x^i} \sum_{\alpha=0}^{\infty} \varepsilon^\alpha \delta u_i^{(\alpha)}(x)\,, 
\end{align}
with the functions inside the summations taken to be periodic in the $x^i$, with period $L^i$. 

We next note that the system of equations \eqref{testokes} (with 
$E^{i}=\zeta^{i}=0$) admit the simple time-independent solution with
$f\,\delta T$, $f\,\delta \mu$ both constant and $\delta u_i=0$. Indeed, from \eqref{fTatmu} this corresponds to
simply perturbing the parameters of the thermal equilibrium configuration. 
The diffusive modes are constructed as a perturbation of this time-independent solution  by
using the expansions \eqref{eq:hydro_fluid_exp}, \eqref{eq:hydro_fluid_exp2} and taking
\begin{align}
f\,\delta T^{(0)}=constant,\qquad f\,\delta \mu^{(0)}=constant,\qquad \delta u^{(0)}_i=0\,,
\end{align}
as the zeroth order solution. We immediately see that the associated expansion for $J^{i}$ and $Q^i$ can be written as 
\begin{align}\label{eq:hydro_fluid_exp_j}
J^{i}=e^{-i\omega t}e^{i \varepsilon k_i x^i} \sum_{\alpha=1}^{\infty} \varepsilon^\alpha J^{i(\alpha)}(x)\,,
\qquad
Q^{i}=e^{-i\omega t}e^{i \varepsilon k_i x^i} \sum_{\alpha=1}^{\infty} \varepsilon^\alpha Q^{i(\alpha)}(x)\,.
\end{align}

At leading order in $\varepsilon$, the first two equations of \eqref{testokes} then read
\begin{align}\label{testokes12o1}
-i\omega^{(1)}\xi_0\delta T^{(0)} -i\omega^{(1)}\chi_0\delta\mu^{(0)} + \nabla_{i}\left(fJ^{i(1)}\right)&=0\,,\notag\\
-i\omega^{(1)}c_{\mu0}f\delta T^{(0)} -i\omega^{(1)}T_{0}\xi_0 f\delta\mu^{(0)} + \nabla_{i}\left(fQ^{i(1)}\right)&=0\,.
\end{align}
Integrating equations \eqref{testokes12o1} over a period we obtain
\begin{align}\label{pin}
i\omega^{(1)}\oint\sqrt{h}\left(\xi_0\,\delta T^{(0)}+\chi_0\, \delta\mu^{(0)} \right)&=0\,,\nn
i\omega^{(1)}\oint\sqrt{h}f\left(c_{\mu0}\,\delta T^{(0)}+T_{0}\xi_0\, \delta\mu^{(0)} \right)&=0\,.
\end{align}
Assuming thermodynamically stable matter, the matrix of static susceptibilities, whose components appear in \eqref{pin}, is positive definite and these equations can only be satisfied by setting $\omega^{(1)}=0$. The leading order system \eqref{testokes} then becomes
\begin{align}\label{testokeso1}
\nabla_{i}\left(fJ^{i(1)}\right)=0\,,\qquad
\nabla_{i}\left(fQ^{i(1)}\right)=0\,,\notag\\
-2\nabla^{i}\left(\eta_{0}\nabla_{(i}\left(f\delta u_{j)}^{(1)}\right)\right) + \nabla_{j}\left(\left(\frac{2\eta_{0}}{(d-1)}-\zeta_{b0}\right)\nabla_{k}\left(f\delta u^{k(1)}\right)\right)&=\notag\\
-i\rho_{0}k_{j}f\delta \mu^{(0)} -\rho_{0}\nabla_{j}\left(f\delta\mu^{(1)}\right) -is_{0}k_{j}f\delta T^{(0)} -s_{0}\nabla_{j}&\left(f\delta T^{(1)}\right)\,.
\end{align}
with
\begin{align}\label{constlin1}
J^{i(1)}&=\rho_{0}\delta u^{i(1)} + \sigma_{Q0}f^{-1}\left[-\nabla^{i}\left(f\delta\mu^{(1)}\right)\right] -\sigma_{Q0}\mu_{0}\left[-(fT_{0})^{-1}\nabla^{i}\left(f\delta T^{(1)}\right)\right]\,,\nn
Q^{i(1)}&=f(P_{0}+\epsilon_{0})\delta u^{i(1)}-f\mu_{0}J^{i(1)}\,.
\end{align}
Notice that this system is equivalent to the system of equations \eqref{testokeso0} that appeared for the calculation of the
thermoelectric 
DC conductivity if we identify $\bar E_{i}\leftrightarrow -ik_{i}f\delta\mu^{(0)}$, $\bar \zeta_{i}\leftrightarrow -ik_{j}T_{0}^{-1}\delta T^{(0)}$
and note that the quantities on the right hand sides of these expressions are indeed constant.
Thus, we can express the heat current fluxes $\bar J^{i(1)}$ and $\bar Q^{i(1)}$ in terms of $-ik_{i}f\delta\mu^{(0)}$, $-ik_{j}T_{0}^{-1}\delta T^{(0)}$ using the thermoelectric DC conductivity matrix given in \eqref{dcmat} to get
\begin{align}\label{fofl}
\bar J^{i(1)}&\equiv \oint\sqrt{h} fJ^{i(1)}=-i\sigma^{ij}_{DC}k_{j}\,f\delta\mu^{(0)}-i\alpha^{ij}_{DC}k_{j}\,f\delta T^{(0)}\,,\nn
\bar Q^{i(1)}&\equiv \oint\sqrt{h} fQ^{i(1)}=-i\bar T_0\alpha^{ij}_{DC}k_{j}\,f\delta\mu^{(0)}-i\bar{\kappa}^{ij}_{DC}k_{j}\,f\delta T^{(0)}\,.
\end{align}

Continuing the expansion, we next examine the first two equations of \eqref{testokes} at second order in $\varepsilon$ to find
\begin{align}\label{hydroeqslast}
-i\omega^{(2)}\xi_0\delta T^{(0)} -i\omega^{(2)}\chi_0\delta\mu^{(0)} +ik_{i}fJ^{i(1)} + \nabla_{i}\left(fJ^{i(2)}\right)&=0\,,\notag\\
-i\omega^{(2)}c_{\mu0}f\delta T^{(0)} -i\omega^{(2)}T_{0}\xi_0 f\delta\mu^{(0)} +ik_{i}fQ^{i(1)} + \nabla_{i}\left(fQ^{i(2)}\right)&=0\,.
\end{align}
Integrating these two equations over a period, substituting the expression for the DC conductivity and using \eqref{totsus},\eqref{totsus2}
we now deduce
\begin{align}\label{impteqs}
i\omega^{(2)}\left(\frac{\partial \rho_{tot}}{\partial \bar T_0}f\delta T^{(0)}+\frac{\partial \rho_{tot}}{\partial \bar \mu_0}f\delta\mu^{(0)}\right)-\alpha^{ij}_{DC}k_{i}k_{j}\,f\delta T^{(0)}-\sigma^{ij}_{DC}k_{i}k_{j}\,f\delta\mu^{(0)}=&0\,,\nn
i\omega^{(2)}\bar T_0\left(\frac{\partial s_{tot}}{\partial \bar T_0}f\delta T^{(0)}+\frac{\partial s_{tot}}{\partial \bar \mu_0}
f\delta\mu^{(0)}\right)-\bar{\kappa}^{ij}_{DC}k_{i}k_{j}\,f\delta T^{(0)}-\bar T_0\alpha^{ij}_{DC}k_{i}k_{j}\,f\delta\mu^{(0)}=&0\,.
\end{align}
Writing this in matrix form as  
\begin{align}
\mathbb{M}\, \left( \begin{array}{c}
f\delta T^{(0)}\\
f\delta\mu^{(0)}
\end{array} \right)=0\,,
\end{align}
we have $\det(\mathbb{M})=0$.  This gives rises to a quadratic equation for $i\omega^{(2)}$ which has two solutions,
$i\omega^{(2)}_{\pm}$, which give the leading order dispersion relations for the diffusion modes that we are after.

To write $i\omega^{(2)}_{\pm}$ in a compact way we first define the scalar quantities depending on the DC conductivities that are quadratic in the
wave numbers $k^i$:
\begin{align}
\bar{\kappa}(k)&\equiv\bar{\kappa}^{ij}_{DC}k_{i}k_{j}\,, \qquad \alpha(k)\equiv\alpha^{ij}_{DC}k_{i}k_{j}\,,\quad \sigma(k)\equiv \sigma^{ij}_{DC}k_{i}k_{j}\,,
\end{align}
as well as
\begin{align}\label{kapksq}
\kappa(k)&\equiv\bar{\kappa}(k)-\frac{\alpha(k)^{2}\bar{T}_{0}}{\sigma(k)}\,.
\end{align}
Recall that $\kappa^{ij}_{DC}\equiv \bar\kappa^{ij}_{DC}-\bar T_0(\bar \alpha_{DC}\cdot\sigma_{DC}^{-1}\cdot\alpha_{DC})^{ij}$ is the DC thermal conductivity for zero electric current and in general $\kappa(k)\ne \kappa^{ij}_{DC}k^i k^j$. 
We also define the following susceptibilities:
\begin{align}\label{susdefs}
X&=\frac{\partial \rho_{tot}}{\partial \bar \mu_0}\,, \qquad \Xi=\frac{\partial s_{tot}}{\partial \bar \mu_0}=\frac{\partial \rho_{tot}}{\partial \bar T_0},\qquad C_{\rho}=\oint\sqrt{h}c_{\mu0}-\frac{\bar{T}_{0}\Xi^{2}}{X} \,.
\end{align}
Note that if we consider the susceptibility $c_\rho=T(\partial s/\partial T)_\rho=c_\mu-\frac{T\xi^2}{\chi}$, in general 
$C_\rho\ne \oint\sqrt{h}c_{\rho0}$. 
Using these definitions, we then find that
\begin{align}\label{finres}
i\omega^{(2)}_{+}\, i\omega^{(2)}_{-}&=\frac{\kappa(k)}{C_{\rho}}\frac{\sigma(k)}{X}\,,\nn
i\omega^{(2)}_{+}+ i\omega^{(2)}_{-}&=\frac{\kappa(k)}{C_{\rho}}+ \frac{\sigma(k)}{X}+\frac{\bar{T}_{0}\,\left(X\,\alpha(k)-\Xi\,\sigma(k) \right)^2}{C_{\rho}X^{2}\sigma(k)}\,.
\end{align}
This is the main result of this section and it should be compared with the general result given in 
\eqref{geneinrel},\eqref{geneinrel2}
that we obtained in the previous section.

A number of comments are in order. Firstly, for relativistic hydrodynamics without a $U(1)$ current, there is just a single energy diffusion mode.
In this case, the leading order dispersion relation is given by
\begin{align}\label{heatdr}
i\omega^{(2)}
=\frac{{\kappa}^{ij}_{DC}k_{i}k_{j}}{\bar T_0\frac{\partial s_{tot}}{\partial \bar T_0}}\,.
\end{align}
This result should be compared with \eqref{feinr}.
Similarly, we can also consider charge neutral backgrounds which have $\Xi=\alpha_{DC}^{ij}=0$ and then the equations 
\eqref{impteqs} decouple. In particular we find a charge diffusion mode with 
leading order dispersion relation given by
\begin{align}
i\omega^{(2)}
=\frac{{\sigma}^{ij}_{DC}k_{i}k_{j}}{\frac{\partial \rho_{tot}}{\partial \bar \mu_0}}\,.
\end{align}

Our next comment concerns perturbative lattices. By definition a perturbative lattice is one in which the metric and gauge field
deformations have a perturbatively small amplitude. In this case the spatial momentum dissipation is weak.
Using the memory matrix formalism \cite{Mahajan:2013cja} or holography \cite{Donos:2015gia} we have
\begin{align}
\bar\kappa_{DC}^{ij}=4\pi s_0T_0 L^{-1}_{ij},\quad 
\alpha_{DC}^{ij}=4\pi \rho_0 L^{-1}_{ij},\quad 
\sigma_{DC}^{ij}=4\pi s_0^{-1}\rho_0^2 L^{-1}_{ij}\,.
\end{align}
Here the matrix $L_{ij}$ incorporates the leading order dissipation and $L_{ij}\to 0$ when translation invariance is retained.
While all of these DC conductivities are large, $\kappa_{DC}^{ij}$ and also $\kappa$ in \eqref{kapksq} are parametrically smaller
as pointed out in \cite{Donos:2014cya,Banks:2015wha}.
Thus, from \eqref{finres} we deduce that one of the frequencies will be proportional to $L^{-1}$ while the other will be parametrically smaller.

\subsubsection{Reduced hydrodynamics}
When translations are broken, it should also be possible to construct a `reduced' hydrodynamical description that just involves
the conserved charges i.e. the heat and the $U(1)$ charge. 
At the level of linear response, this can be done, in principle, 
by solving for $\delta u_i^{(n)}$ order by order in the equations \eqref{testokes}, to, effectively,  
get a set of linear equations for the variables $\delta T$ and $\delta \mu$ and highly non-local in terms of the background
metric and gauge-field. We will not carry out this in any detail here, but instead highlight
some interesting features of the leading order terms that would arise. In particular, we will be able to derive a set of 
reduced hydrodynamical equations, at the level of linear response, that generalise those discussed in \cite{Hartnoll:2014lpa}.

We begin with the on-shell expressions for the currents in the $\varepsilon$ expansion given in \eqref{eq:hydro_fluid_exp_j}. Focussing
on the $U(1)$ current for the moment, we recall that at each order 
$\sqrt{h} f J^{i(n)}$ are periodic functions of the $x^i$.
We have seen that at leading order they are determined by the system of linear equations given in \eqref{testokeso1},
 which is equivalent to the system of equations \eqref{testokeso0} that appeared for the calculation of the
DC conductivity if we identify $\bar E_{i}\leftrightarrow -ik_{i}f\delta\mu^{(0)}$, $\bar \zeta_{i}\leftrightarrow -ik_{j}T_{0}^{-1}\delta T^{(0)}$. We can therefore write $\sqrt{h} f J^{i(1)}$ linearly in terms of $f\delta\mu^{(0)}$, $f\delta T^{(0)}$
as a sum of a constant flux, expressed in terms of the DC conductivity matrix, and a term
which is co-closed and has vanishing zero mode (a periodic magnetisation current). 
Thus, we can write for the full current
\begin{align}
\sqrt{h} f J^{i}&=e^{-i\omega t}e^{i \varepsilon k_i x^i}
\varepsilon\left[(\sigma^{ij}_{DC}+\partial_kS^{kij})(-ik_{j}\,f\delta\mu^{(0)})
+(\alpha^{ij}_{DC}+\partial_k A^{kij})(-ik_jf\delta T^{(0)})+\mathcal{O}(\varepsilon) \right]\,,
\end{align}
where $S^{kij}=-S^{ikj}$, $A^{kij}=-A^{ikj}$ and both are periodic functions of the spatial coordinates.
We can also obtain a similar expression for the heat current and we can write both of them in the following suggestive form
\begin{align}\label{exs}
\sqrt{h} f J^{i}&=-(\sigma_{DC}^{ij} +\partial_kS^{kij}  )\nabla_{j}\delta\hat\mu -(\alpha_{DC}^{ij}+\partial_k A^{kij})\nabla_{j}\delta\hat T  + \dots\,,\nn
\sqrt{h} f Q^{i}&=-\bar T_0(\alpha_{DC}^{ij}+\partial_k A^{kij})\nabla_j\delta\hat\mu-(\bar{\kappa}^{ij}_{DC}+\partial_k K^{kij})\nabla_j\delta\hat T + \dots\,,
\end{align}
where $\delta\hat\mu\equiv e^{-i\omega t} e^{i \varepsilon k_i x^i}f\delta\mu^{(0)}$, 
$\delta\hat T\equiv e^{-i\omega t} e^{i \varepsilon k_i x^i}f\delta T^{(0)}$ and $K^{kij}=-K^{ikj}$. In these on-shell expressions $\omega$ is fixed as an expansion in $\varepsilon$ in terms of $k_i$ and the background quantities via the dispersion relations.

We next consider analogous expressions for the local charge density and heat density. From \eqref{tjperturbative}
we obtain 
\begin{align}
\sqrt{h}f J^t&=\sqrt{h}\rho_0+e^{-i\omega t}e^{i \varepsilon k_i x^i}\sqrt{h}\left[\xi_0\delta T^{(0)}+
\chi_0\delta \mu^{(0)}+\mathcal{O}(\varepsilon)\right]\,,\nn
\sqrt{h}f Q^t&=\sqrt{h}f(\epsilon_0-\mu_0\rho_0)+e^{-i\omega t}e^{i \varepsilon k_i x^i}\sqrt{h}f\left[c_{\mu0}\delta T^{(0)}+
T_0\xi_0\delta \mu^{(0)}+\mathcal{O}(\varepsilon)\right]\,,
\end{align}
where $\sqrt{h}\rho_0$ and $\sqrt{h}f(\epsilon_0-\mu_0\rho_0)$ are the local charge densities in equilibrium. Hence, for the perturbation
we can write
\begin{align}\label{chgesaux}
\delta[\sqrt{h}f J^t]&= \sqrt{h}f^{-1}\xi_0\delta\hat T+\sqrt{h}f^{-1}\chi_0\delta\hat\mu+\dots\,,\nn
\delta[\sqrt{h}f Q^t]&= \sqrt{h}c_{\mu0}\delta\hat T+T_0\sqrt{h}\xi_0\delta\hat\mu+\dots\,.
\end{align}

At this stage, from these on-shell expressions, 
we now can see the leading order structure of an off-shell reduced hydrodynamics. Specifically,
if we take \eqref{chgesaux} to be expressions for the local charge densities and
\eqref{exs} to be the associated constitutive relations for the currents, the continuity equations $\nabla_\mu J^\mu=\nabla_\mu Q^\mu=0$
at order $\varepsilon^2$ will lead to the same diffusive solutions that we had above with
exactly the same dispersion relations for the diffusion modes. 
In particular, the magnetisation currents in \eqref{exs} do not play a role in this specific calculation. 
It is also worth emphasising that in this reduced hydrodynamics, the variables $\delta\hat T$, $\delta\hat\mu$ need not
be periodic functions and indeed they are not in the diffusive solutions.

We can now compare these results with the hydrodynamics described in 
the ``Methods" section of \cite{Hartnoll:2014lpa}, highlighting several differences.
Firstly, the constitutive relations for the local currents given in \cite{Hartnoll:2014lpa} were declared to be given in terms of the DC conductivity, whereas here we have derived them from the underlying relativistic hydrodynamics. Secondly,
the possibility of the terms involving $S^{kij}$, $A^{kij}$, $K^{kij}$ was not considered in \cite{Hartnoll:2014lpa}. 
Finally, the expression for the local charge densities in \cite{Hartnoll:2014lpa} were not of the form \eqref{chgesaux}. 
To make a connection we note that using \eqref{totsus}, \eqref{totsus2}
we can rewrite \eqref{chgesaux} in the form
\begin{align}\label{chgesaux2}
\delta[\sqrt{h}f J^t]&= \left(\frac{\partial \rho_{tot}}{\partial \bar T_0}+\dots\right)\delta\hat T
+\left(\frac{\partial \rho_{tot}}{\partial \bar \mu_0}+\dots \right)\delta\hat\mu+\dots\,,\nn
\delta[\sqrt{h}f Q^t]&= \left(  \bar T_0\frac{\partial s_{tot}}{\partial \bar T_0} +\dots \right)\delta\hat T+\left(  \bar T_0\frac{\partial s_{tot}}{\partial \bar \mu_0} +\dots \right)\delta\hat\mu+\dots\,.
\end{align}
where in the bracketed terms we have just written the constant zero mode part of the relevant term.
The expressions \eqref{chgesaux2} are what were considered in \cite{Hartnoll:2014lpa}; while the neglected higher Fourier modes will not affect the
calculation of the dispersion relations for the diffusive modes, they are the same order in the $\varepsilon$ expansion with the zero modes and they should be included as they will affect other calculations.

\subsubsection{Green's functions}
Within the context of relativistic hydrodynamics, the leading order solutions for the charge density and the currents are given 
in the previous subsection. It is possible to relate these expressions to the retarded Green's functions. At a first pass this seems problematic as the diffusive solutions are source free solutions and yet to extract Green's functions we need to relate a response to a source. 

This puzzle can be resolved by 
the following trick. We view the solutions as having arisen after adiabatically switching on sources for the charge density in the far past, switching them off at time $t=0$ and then comparing the solutions for $t>0$ in the long wavelength limit. As this is somewhat technical
we have explained how this can be achieved, as well as presenting some results of general validity, in appendix \ref{appa}. For simplicity, 
we will carry out the analysis just for the case when there is only a single current present, which is the heat current. 
Hence, for convenience we present the perturbed part of the diffusive solution in this case here:
\begin{align}\label{bulksumsol}
\delta[\sqrt{h}f Q^t]&= e^{-i\omega t}e^{i \varepsilon k_i x^i}\left[ \sqrt{h}c_{\mu0}f\delta T^{(0)}+\mathcal{O}(\varepsilon)\right]\,,\nn
\sqrt{h} f Q^{i}&=e^{-i\omega t}e^{i \varepsilon k_i x^i}\varepsilon \left[(\bar{\kappa}^{ij}_{DC}+\partial_k K^{kij})(-ik_j)f\delta T^{(0)}+\mathcal{O}(\varepsilon)\right]\,,
\end{align}
with $i\omega
=\frac{{\kappa}^{ij}_{DC}k_{i}k_{j}}{\bar T_0\frac{\partial s_{tot}}{\partial \bar T_0}}$. We also
recall that $f\delta T^{(0)}$ is constant and $\sqrt{h}c_{\mu0}$ is a local susceptibility whose constant zero mode piece is
$\bar T_0\frac{\partial s_{tot}}{\partial \bar T_0}$.

\section{Final comments}
In this paper we have made a general study of the hydrodynamical diffusion modes associated with conserved charges that arise in inhomogeneous media with a lattice symmetry. When the DC conductivities are finite, we showed that there are diffusive modes with dispersion relations that are determined by the DC conductivities and certain thermodynamical susceptibilities. This constitutes a generalised Einstein relation for inhomogeneous media.
We also illustrated the general results, obtained by an analysis of retarded Greens functions, by considering the specific context of relativistic hydrodynamics.
For simplicity, here we have focused on systems that are invariant under time reversal. However, it should be straightforward to generalise to the non-static case, after identifying suitably defined transport currents as in \cite{PhysRevB.55.2344,Hartnoll:2007ih,Blake:2015ina,Donos:2015bxe,Donos:2017oym}.

In \cite{Banks:2016wdh}, 
for a general conformal field theory on a curved manifold with a metric of the form \eqref{bfield} with $f=1$, $h_{ij}=\Phi\delta_{ij}$ and
$\Phi$ a periodic function, the relativistic hydrodynamic equations (with vanishing $U(1)$ fields) were solved for the local temperature and 
heat current, at the level of linear response, after applying a DC thermal gradient $\tilde\zeta_i$. 
In particular, it was shown that thermal backflow can occur whereby the heat current is locally flowing in
the opposite direction to the DC source. 
These results can be recast in terms of the diffusion results of this paper. Let $\omega^{(2)}$ be the leading order dispersion relation as
in \eqref{heatdr}. Then, focussing on real variables, we have leading order diffusing solutions with
$\delta T=e^{-\varepsilon^2\omega^{(2)} t}\cos(\varepsilon k_i x^i)(\delta T^{(0)}+\varepsilon \delta T^{(1)}+\mathcal{O}(\varepsilon^2))$,
and the local heat current given by $\delta Q^i=e^{-\varepsilon^2\omega^{(2)} t}\sin(\varepsilon k_i x^i)\varepsilon(\delta Q^{i(1)}+\mathcal{O}(\varepsilon))$, where
$\delta T^{(1)}$ and $\delta Q^{i(1)}$ are the local temperature and heat current obtained in \cite{Banks:2016wdh} for a DC thermal gradient given by $\tilde\zeta_i=k_i\delta T^{(0)}$.
We can consider these solutions as having been adiabatically prepared in an initial state 
at $t=0$ (say) and then diffusing. The solution shows that in each individual spatial period
there is an elaborate local structure, which includes thermal backflow, with an overall damping of the current in time. 

The existence of the same backflow current patterns that emerge in the steady state set-up provides a non-trivial test of the validity of hydrodynamics for certain strongly correlated systems of electrons for which backflows have been observed.
Finally, we note that the initial conditions at $t=0$ that we are considering, arising from the construction of specific long wavelength diffusion modes,
might seem fine tuned. However, as long as short wavelength modes die out faster in time, the diffusive modes will capture the universal 
late time behaviour for generic initial conditions. For systems with light, spatially modulated modes, this will be case provided we examine long enough wavelengths.

The general results of this paper should also manifest themselves within the context of holography. In particular, it should be possible
to obtain the Einstein relations in terms of the DC conductivities and the thermodynamic susceptibilities. It is now understood
how, in general, the thermoelectric DC conductivity of the boundary field theory, when finite, can be obtained in terms of data on the black hole horizon \cite{Donos:2015gia,Banks:2015wha,Donos:2015bxe}. 
Thus, providing one can obtain
the susceptibilities in terms of horizon data, one should also be able to extract the Einstein relations. This will be explored in \cite{Donos:2017ihe}. This line of investigation could also make contact with the recent work on relating diffusion to a characteristic velocity extracted from the black hole horizon, related to out of time ordered correlators
\cite{Blake:2016wvh,Blake:2016sud} and \cite{Lucas:2016yfl,Ling:2016ibq,Blake:2016jnn,Davison:2016ngz,Baggioli:2016pia,Wu:2017mdl,Kim:2017dgz,Baggioli:2017ojd,Blake:2017qgd,Wu:2017exh}.

\section*{Acknowledgements}
We thank Tom Griffin and Luis Melgar for helpful discussions.
The work of JPG is supported by the European Research Council under the European Union's Seventh Framework Programme (FP7/2007-2013), ERC Grant agreement ADG 339140, STFC grant ST/L00044X/1, EPSRC grant EP/K034456/1, as a KIAS Scholar and as a Visiting Fellow at the Perimeter Institute. VZ is supported by a Faculty of Science Durham Doctoral Scholarship.

\appendix
 
\section{General results}\label{genrsa}
Here we present some general results for Green's functions involving a single conserved current density operator $J^\mu$ satisfying
the continuity equation $\partial_\mu J^\mu=0$. We will present results for $G_{J^{\mu}J^\nu}(\omega,\mathbf{k},\mathbf{k}^{\prime})$;
using the crystallographic decomposition \eqref{eq:g_cystal} we can easily extract
analogous results for the $G^{(\{n_{j}\})}_{J^{\mu}J^\nu}(\omega,\mathbf{k})$.

From \eqref{eq:g_fourier} the current conservation condition $\partial_\mu J^\mu=0$ implies
\begin{align}\label{eq:fourier_ward1}
-i\omega\, G_{J^{t}B}(\omega,\mathbf{k},\mathbf{k}^{\prime})+i\mathbf{k}_{i}G_{J^{i}B}(\omega,\mathbf{k},\mathbf{k}^{\prime})=0\,,
\end{align}
for any operator ${B}$, whose equal time commutator with $J^t$ vanishes. From \eqref{eq:fourier_ward1} we have
\begin{align}
-i\omega\, G_{J^{t}J^{t}}(\omega,\mathbf{k},\mathbf{k}^{\prime})+i\mathbf{k}_{i}G_{J^{i}J^{t}}(\omega,\mathbf{k},\mathbf{k}^{\prime})=&0\,,\notag\\
-i\omega\, G_{J^{t}J^{j}}(\omega,\mathbf{k},\mathbf{k}^{\prime})+i\mathbf{k}_{i}G_{J^{i}J^{j}}(\omega,\mathbf{k},\mathbf{k}^{\prime})=&0\label{eq:j_ide1}\,.
\end{align}

We next consider the time reversal invariance conditions \eqref{trevcon}.
Since $\epsilon_{J^t}=+1$ and $\epsilon_{J^i}=-1$, we obtain
\begin{align}\label{eq:t_reversal}
G_{J^{i}J^{t}}(\omega,\mathbf{k},\mathbf{k}^{\prime})&=-G_{J^{t}J^{i}}(\omega,-\mathbf{k}^{\prime},-\mathbf{k})\,,\nn
G_{J^{i}J^{j}}(\omega,\mathbf{k},\mathbf{k}^{\prime})&=G_{J^{j}J^{i}}(\omega,-\mathbf{k}^{\prime},-\mathbf{k})\,.
\end{align}
Combing \eqref{eq:t_reversal} with \eqref{eq:j_ide1} we therefore have 
\begin{align}\label{krone1}
\mathbf{k}_{i}\mathbf{k}^{\prime}_{j}\,G_{J^{i}J^{j}}(\omega,\mathbf{k},\mathbf{k}^{\prime})&=-(i\omega)^2\,G_{J^{t}J^{t}}(\omega,\mathbf{k},\mathbf{k}^{\prime})\,,\nn
i\mathbf{k}^{\prime}_{j}\,G_{J^{i}J^{j}}(\omega,\mathbf{k},\mathbf{k}^{\prime})&=(i\omega)\,G_{J^{i}J^{t}}(\omega,\mathbf{k},\mathbf{k}^{\prime})\,.
\end{align}

Define the static susceptibility
\begin{align}\label{defcsus2}
\lim_{\omega\to 0+i0}G_{J^{t}J^{t}}(\omega,\mathbf{k},\mathbf{k}^{\prime})\equiv-\chi_{J^{t}J^{t}}(\mathbf{k,\mathbf{k}^{\prime}})\,.
\end{align}
We see that \eqref{eq:j_ide1} and \eqref{krone1} imply
\begin{align}\label{ccsusc}
\mathbf{k}_{i}\chi_{J^{i}J^{t}}(\mathbf{k,\mathbf{k}^{\prime}})&=0\,,\nn
\mathbf{k}_{i}\mathbf{k}^{\prime}_{j}\chi_{J^{i}J^{j}}(\mathbf{k,\mathbf{k}^{\prime}})&=0\,.
\end{align}
Note that the sign in \eqref{defcsus2} is fixed as follows. From \eqref{eq:response}, for a time-independent source for the charge density $\delta h_{J^t}(\mathbf{x})$, we have
\begin{align}\label{dfol}
 \delta \langle {J}^{t}\rangle (t,\mathbf{k})= \frac{1}{(2\pi)^d}\int d\mathbf{k}'G_{J^{t}J^{t}}(\omega=0,\mathbf{k},\mathbf{k}')\,\delta h_{J^{t}}(\mathbf{k}')\,.
\end{align}
On the other hand from \eqref{delaich} $\delta H=(2\pi)^{-d}\int d\mathbf{k}\delta h_{J^t}(-\mathbf{k})\delta J^t(\mathbf{k})$
and so we identify the perturbed chemical potential, $\delta\mu(\mathbf{k})$, as $\delta\mu(\mathbf{k})=- \delta h_{J^{t}}(\mathbf{k})$.
Since the static susceptibility $\chi_{J^{t}J^{t}}$ is defined by varying the charge density
with respect to the chemical potential we get the sign as in \eqref{defcsus2}.

 \section{Linear response from a prepared source}\label{appa}
 We consider a perturbative deformation of the Hamiltonian as in \eqref{delaich}, with a prepared source that is switched off at $t=0$, 
 given by
 \begin{align}\label{sceapp} 
 h_{B}(t,\mathbf{x})=\begin{cases}
 e^{\varepsilon_{t}\,t+i\,\mathbf{k}_{s}\,\mathbf{x}}\,\delta h_{B},& t\le 0\\
0& t>0
 \end{cases}\,,
 \end{align}
 with $\varepsilon_{t}>0$. This source contains a single spatial Fourier mode and we will be interested in taking the adiabatic limit $\varepsilon_{t}\to 0^{+}$.
  
 The time dependent expectation value of an operator ${A}$ is given by the retarded Green's function
 as in \eqref{eq:response}.
Thus, at $t=0$, when the sources are switched off, we have 
 \begin{align}\label{tequalz}
 \delta \langle {A}\rangle (t=0,\mathbf{x})&=\int\,dt^{\prime}\,d\mathbf{x}^{\prime}\,G_{AB}(-t^{\prime},\mathbf{x},\mathbf{x}^{\prime})\,\delta h_{B}(t^{\prime},\mathbf{x}^{\prime})\,,\nn
 &=\int\,dt^{\prime}\,d\mathbf{x}^{\prime}\,G_{AB}(t^{\prime},\mathbf{x},\mathbf{x}^{\prime})\, e^{-\varepsilon_{t}t^{\prime}+i\,\mathbf{k}_{s}\mathbf{x}^{\prime}}\,\delta h_{B}\,,\nn
 &=G_{AB}(i\,\varepsilon_{t},\mathbf{x},\mathbf{k}_{s})\,\delta h_{B}\,.
 \end{align} 
In the $\varepsilon_{t}\to 0^{+}$ limit, after a Fourier transform, we have 
 \begin{align}\label{susapp}
 \delta \langle {A}\rangle (t=0,\mathbf{k})= -\chi_{AB}(\mathbf{k},\mathbf{k}_{s})\,\delta h_{B}\,,
 \end{align}
where $\chi_{AB}(\mathbf{k},\mathbf{k}')\equiv -\lim_{\omega\to 0+i0}G_{AB}(\omega,\mathbf{k},\mathbf{k}')$.
Also, after a Fourier transform of the source \eqref{sceapp}, for any $t>0$ we deduce that
\begin{align}
\delta \langle {A}\rangle (t,\mathbf{x})=\frac{1}{2\pi}\int_{-\infty}^{+\infty}\,d\omega\,\frac{1}{\varepsilon_{t}+i\,\omega}e^{-i\omega\,t}G_{AB}(\omega,\mathbf{x},\mathbf{k}_{s})\,\delta h_{B}\,.
\end{align}
Taking a Laplace transform in time we get
\begin{align}
\delta \langle{A}\rangle (z,\mathbf{x})&\equiv \int_{0}^{+\infty}dt\,\delta \langle {A}\rangle (t,\mathbf{x})\,e^{i\,z t}\,,\nn
&=-\frac{1}{2\pi}\,\int_{-\infty}^{+\infty}\,d\omega\,\frac{1}{\omega-i\,\varepsilon_{t}}\frac{1}{\omega-z}\,G_{AB}(\omega,\mathbf{x},\mathbf{k}_{s})\,\delta h_{B}\,,
\end{align}
with, necessarily, $\mathrm{Im}\, z>0$ in order for the integrals to converge. Performing a contour integral on the above expression by closing it in the upper half plane and 
assuming that the Green's function vanishes fast enough for large $\omega$, we just pick up contributions from the poles at $\omega=i\,\varepsilon_{t}$ and $\omega=z$ to obtain
\begin{align}\label{thisonelim}
\delta \langle {A}\rangle (z,\mathbf{x})=-\frac{i}{i\,\varepsilon_{t}-z} G_{AB}(i\,\varepsilon_{t},\mathbf{x},\mathbf{k}_{s})\,\delta h_{B}-\frac{i}{z-i\,\varepsilon_{t}}G_{AB}(z,\mathbf{x},\mathbf{k}_{s})\,\delta h_{B}\,.
\end{align}
Thus, in the $\varepsilon_{t}\to 0^{+}$ limit we conclude that the spatial Fourier transform is given by \eqref{sceapp}:
\begin{align}
\delta \langle {A}\rangle (z,\mathbf{k})
&=\frac{1}{i\,z}\left(G_{AB}(z,\mathbf{k},\mathbf{k}_{s})+ \chi_{AB}(\mathbf{k},\mathbf{k}_{s})\right)\,\delta h_{B}\,.
\end{align}
Using \eqref{susapp} we now obtain the following solution to the initial value problem that is sourced by \eqref{sceapp} in
the $\varepsilon_{t}\to 0^{+}$ limit:
\begin{align}\label{lastdelaap}
\delta \langle {A}\rangle (z,\mathbf{k})
&=-\frac{1}{i\,z}\left(G_{AB}(z,\mathbf{k},\mathbf{k}_{s})\chi^{-1}_{BC}(\mathbf{k},\mathbf{k}_{s})+\delta_{AC} \right) \delta \langle {C}\rangle (t=0,\mathbf{k})\,.
\end{align}

\subsection{Conserved current}

Let us now apply some of these results to conserved currents. For simplicity we just consider the case of a single conserved current and assume that
there is a single diffusion pole.
We will assume that the source \eqref{sceapp} is a source just for the charge density operator $J^t$. In particular at $t=0$ we write the source as $e^{i\mathbf{k}_s\mathbf{x}}\delta h^{(0)}_{J^t}$, with constant $\delta h^{(0)}_{J^t}$. 
We take the limit $\varepsilon_t\to 0$ and then consider $\mathbf{k}_s\to 0$.

From \eqref{thisonelim}, the time dependence of the charge density for $t>0$ is fixed by the Laplace transformed quantity
\begin{align}\label{genie}
 \delta \langle {J^t}\rangle (z,\mathbf{x})
 &=e^{i\,\mathbf{k}_s\mathbf{x}}\,
\sum_{\{n_{j}\}}e^{i n_{j}\mathbf{k}^{j}_{L}\mathbf{x}}  \frac{1}{iz}\left[G^{(\{n_{j}\})}_{J^tJ^t}(z,\mathbf{k}_s)+
\chi^{(\{n_{j}\})}_{J^tJ^t}(\mathbf{k}_s)
\right]\frac{1}{(2\pi)^{d}}\delta h^{(0)}_{J^t}\,,
\end{align}
where 
$\chi^{(\{n_{j}\})}_{J^tJ^t}(\mathbf{k}_s)=-\lim_{\omega\to 0+i0}G^{(\{n_{j}\})}_{J^tJ^t}(\omega,\mathbf{k}_s)$. It is interesting to now examine the
zero mode of the periodic function inside the sum (see \eqref{eq:response3}):
\begin{align}
 \delta \langle {J^t}\rangle^{\{(0)\}} (z)
 &=  \frac{1}{iz}\left[G_{J^tJ^t}(z,\mathbf{k}_s)+
\chi_{J^tJ^t}(\mathbf{k}_s)
\right]\frac{1}{(2\pi)^{d}}\delta h^{(0)}_{J^t}\,,
\end{align}
since we can draw some further general conclusions using the results of section \ref{secsc}. Indeed
after considering $\mathbf{k}_s\to 0$, and recalling the general results \eqref{geenen} and \eqref{ennexp}, we have
\begin{align}
 \delta \langle {J^t}\rangle^{\{(0)\}} (z)
 &=  \frac{-1}{-iz+\mathbf{k}_{si}\mathbf{k}_{sj}\sigma^{ij}(z)\chi(\mathbf{0})^{-1}}\chi(\mathbf{0})\frac{1}{(2\pi)^{d}}\delta h^{(0)}_{J^t}\,.
\end{align}
Taking the inverse Laplace transform and keeping just the time-dependence that is leading order in $\mathbf{k}_s$, we obtain
\begin{align}
 \delta \langle {J^t}\rangle^{\{(0)\}} (t)
 &=  -e^{-i\omega(\mathbf{k}_s) t}\chi(\mathbf{0})\frac{1}{(2\pi)^{d}}\delta h^{(0)}_{J^t}\,.
\end{align}
with $i\omega(\mathbf{k}_s)
={\sigma}^{ij}_{DC}\mathbf{k}_{si}\mathbf{k}_{sj}\chi(\mathbf{0})^{-1}$.

We can now make a comparison with the diffusive solutions given in \eqref{bulksumsol} that we found within the context of relativistic hydrodynamics.  
Recalling that in this appendix, and also in section \ref{green}, we are considering current densities, whereas in section \ref{relhsec} we used current vectors, we therefore should compare the local current $\delta[\sqrt{h}f Q^t(t,\mathbf{x})]$ in \eqref{bulksumsol}
with $\delta \langle {J^t}\rangle (t,\mathbf{x})$. Identifying the constant
source $\frac{1}{(2\pi)^{d}}\delta h^{(0)}_{J^t}$ here with $-f\delta T^{(0)}$ (see the discussion following \eqref{dfol}), after comparing \eqref{bulksumsol} with \eqref{genie} and the above analysis, we conclude that
for these particular solutions we have that for each $\{n_{j}\}$, in the limit that $\mathbf{k}_s\to 0$,
\begin{align}
G^{(\{n_{j}\})}_{J^tJ^t}(\omega,\mathbf{k}_s)\chi^{(\{n_{j}\})}_{J^tJ^t}(\mathbf{k}_s)^{-1}+1
\to  \frac{1}{-i\omega+\mathbf{k}_{si}\mathbf{k}_{sj}S^{\{n_{j}\}ij}(\omega)\chi(\mathbf{0})^{-1}}\,,
\end{align}
with $S^{\{n_{j}\}ij(\omega)}=\sigma_{DC}^{ij}+\mathcal{O}(\omega)$, in order to get the correct time-dependence. 
In particular, all of these modes of the Green's function have the same
diffusion pole at the origin.

We next consider the spatial components of the current. Starting with \eqref{lastdelaap} and using \eqref{krone1}
we can write
\begin{align}
\delta \langle {J^i}\rangle (z,\mathbf{k})=
\left[\frac{1}{(iz)^2}G_{J^iJ^j}(z,\mathbf{k},\mathbf{k}_s)(-i\mathbf{k}_{sj}) +\frac{1}{iz}\chi_{J^iJ^t}(\mathbf{k},\mathbf{k}_s)
\right]\delta h^{(0)}_{J^t}\,.
\end{align}
After a Fourier transform on the spatial coordinates we can therefore write
\begin{align}\label{smthtocomp}
\delta \langle {J^i}\rangle (z,\mathbf{x})
=e^{i\,\mathbf{k}_s\mathbf{x}}\,
\sum_{\{n_{j}\}}e^{i n_{j}\mathbf{k}^{j}_{L}\mathbf{x}}  \left[\frac{1}{(iz)^2}G^{(\{n_{j}\})}_{J^iJ^j}(z,\mathbf{k}_s)(-i\mathbf{k}_{sj})
+\frac{1}{iz}\chi^{(\{n_{j}\})}_{J^iJ^t}(\mathbf{k}_s)
\right]\frac{1}{(2\pi)^{d}}\delta h^{(0)}_{J^t}\,.
\end{align}

Current conservation implies that $\mathbf{k}_{i}\chi_{J^iJ^t}(\mathbf{k},\mathbf{k}_s)=0$ (see \eqref{ccsusc}) but in general
$\chi_{J^iJ^t}(\mathbf{k},\mathbf{k}_s)\ne 0$. However, in the relativistic hydrodynamics in the static background
we do have $\chi_{J^iJ^t}(\mathbf{k},\mathbf{k}_s)=0$ (see the comment below \eqref{tjperturbativeeq}). Thus,
comparing \eqref{smthtocomp} with \eqref{bulksumsol} we deduce that for the relativistic hydrodynamics, as $\mathbf{k}_s\to 0$ we have 
\begin{align}
\frac{1}{(i\omega)^2}G^{(\{n_{j}\})}_{J^iJ^j}(\omega,\mathbf{k}_s)(-i\mathbf{k}_{sj})
\to
 \frac{(\bar{\kappa}^{ij}_{DC}+\partial_k K^{kij})^{(\{n_{j}\})}(-i\mathbf{k}_{sj})}{-i\omega+\mathbf{k}_{si}\mathbf{k}_{sj}\tilde S^{\{n_{j}\}ij}(\omega)\chi(\mathbf{0})^{-1}}\,,
\end{align}
with $\tilde S^{\{n_{j}\}ij(\omega)}=\sigma_{DC}^{ij}+\mathcal{O}(\omega)$ in order to get the correct time-dependence.

A final comment is that if we consider \eqref{susapp} with
$\chi_{J^iJ^t}(\mathbf{k},\mathbf{k}_s)=0$ then we deduce that $ \delta \langle {J^i}\rangle (t=0,\mathbf{x})=0$. 
This seems inconsistent with the $t=0$ limit of the diffusive solution arising from hydrodynamics. The resolution of this puzzle
is that when we take the limit $\varepsilon_t\to 0$ it leads to a discontinuity in the current. The correct thing to do is compare the currents
for $t>0$ as we did above.


\providecommand{\href}[2]{#2}\begingroup\raggedright\endgroup

\end{document}